\pdfoutput=1

\documentclass[aps,pre,preprint,superscriptaddress]{revtex4-1}

\usepackage[caption=false]{subfig}
\usepackage{graphicx}
\usepackage{amsmath}
\usepackage{amssymb}
\usepackage[hidelinks]{hyperref}
\usepackage[normalem]{ulem}

\newcommand*\diff{\mathop{}\!\mathrm{d}}

\usepackage{color}

\begin{document}

\title{Controlling permeation in electrically-deforming liquid crystal network films: a dynamical Landau theory}

\author{Guido L. A. Kusters}
\email[]{g.l.a.kusters@tue.nl}
\affiliation{Department of Applied Physics, Eindhoven University of Technology, The Netherlands}

\author{Nicholas B. Tito}
\affiliation{Electric Ant Lab, Amsterdam, The Netherlands}

\author{Cornelis Storm}
\affiliation{Department of Applied Physics, Eindhoven University of Technology, The Netherlands}
\affiliation{Institute for Complex Molecular Systems, Eindhoven University of Technology, The Netherlands}

\author{Paul van der Schoot}
\affiliation{Department of Applied Physics, Eindhoven University of Technology, The Netherlands}

\date{\today}

\begin{abstract}
Liquid crystal networks exploit the coupling between the responsivity of liquid-crystalline mesogens, e.g., to electric fields, and the (visco)elastic properties of a polymer network. Because of this, these materials have been put forward for a wide array of applications, including responsive surfaces such as artificial skins and membranes. For such applications, the desired functional response must generally be realized under strict geometrical constraints, such as provided by supported thin films. To model such settings, we present a dynamical, spatially-heterogeneous Landau-type theory for electrically-actuated liquid crystal network films. We find that the response of the liquid crystal network permeates the film from top to bottom, and illustrate how this affects the time scale associated with macroscopic deformation. Finally, by linking our model parameters to experimental quantities, we suggest that the permeation rate can be controlled by varying the aspect ratio of the mesogens and their degree of orientational order when cross-linked into the polymer network, for which we predict a single optimum. Our results contribute specifically to the rational design of future applications involving transport or on-demand release of molecular cargo in liquid crystal network films.
% [G. L. A. Kusters, I.P. Verheul, N. B. Tito, P. van der Schoot, and C. Storm, Phys. Rev. E \textbf{102}, 042703 (2020)]
\end{abstract}

\pacs{aaa}

\maketitle

\section{Introduction}\label{sec:Introduction}
Liquid crystal networks, like nematic elastomers, couple the orientational anisotropy of liquid-crystalline mesogens to macroscopic deformations by crosslinking the mesogens into a polymer network \cite{finkelmann1981investigations}. It has been shown that this coupling can be exploited in practical applications: by influencing the orientation of the mesogens, for example through a temperature- or electric-field-induced isotropic-to-nematic phase transition \cite{gramsbergen1986landau,prost1995physics}, deformations are achievable far exceeding those that can be expected based on the constituent elements \cite{warner1996nematic,tajbakhsh2001spontaneous}. Moreover, since the form of the resulting deformation is contingent on the configuration with which the material is synthesized and prepared \cite{ohm2010liquid,white2015programmable}, even ``programming" the desired response into the material, for example for shape-memory materials \cite{rousseau2003shape,liu2007review,burke2010soft,lee2011light,lee2012autonomous,burke2013evolution,kotikian20183d}, haptic feedback \cite{campo2011nano,camargo2011microstamped,camargo2011localised,camargo2012batch,torras2014tactile}, or artificial muscles \cite{wermter2001liquid,ikeda2007photomechanics,sanchez2009photo,sanchez2011liquid,dai2013humidity,schuhladen2014iris,iamsaard2016fluorinated,zeng2017self}, is possible.

Recently, liquid crystal networks have enjoyed particular experimental scrutiny in the context of responsive coatings. These surfaces generally capitalize on the tunable dynamical behavior of the liquid crystal network \cite{smith2008review,heo2015fast}, in combination with their large susceptibility to external stimuli, such as temperature \cite{cao2019temperature,babakhanova2019surface}, electric fields \cite{liu2017protruding,van2019morphing,van2020electroplasticization}, or UV irradiation \cite{white2012light,stumpel2014stimuli}. Topical highlights include the realization of controllable surface topographies \cite{liu2012photo,mcconney2013topography,liu2017protruding,babakhanova2019surface,van2019morphing,van2020electroplasticization}, adaptive adhesion and friction \cite{liu2014self,gelebart2018photoresponsive}, as well as artificial skins and membranes \cite{cao2019temperature,zhan2020localized}.

An important common denominator is that these applications all concern thin films in which the local transport of material or molecular cargo can be expected to play a prominent role. As such, a pressing design challenge is to gain spatial control over the response of the liquid crystal network. Although the spatially-resolved response in various polymeric materials has been studied \cite{stuart2010emerging,cui2015dynamic,cao2019temperature,zhan2020localized}, this has yet to lead to clear experimental guidelines for liquid crystal networks. This is where a theoretical approach could prove valuable, complementing experimental work toward a rational design strategy and minimizing costly trial and error.

To meet the need for this theoretical support, the aim of the current paper is to provide a theoretical framework for the spatially-resolved deformation of liquid crystal networks, which we subsequently use to study how the response permeates the liquid crystal network in a (clamped) thin-film geometry. One approach would be to build upon the so-called ``neo-classical" theory of nematic elastomers, as proposed by Warner, Terentjev and others in their pioneering work \cite{warner1988theory,warner1991elasticity}. Indeed, this framework, based on a combination between the classical theory of elasticity and the Landau-de Gennes theory of liquid crystals, is well-celebrated and has been studied extensively, shedding light on, among other things, the phase transitions and instabilities nematic elastomers exhibit \cite{bladon1993transitions,verwey1996elastic,tajbakhsh2001spontaneous}, their response to electric fields \cite{terentjev1994orientation}, as well as the existence of (quasi)soft modes of deformation \cite{golubovic1989nonlinear,warner1994soft,olmsted1994rotational,verwey1995soft,martinoty2004mechanical,terentjev2004commentary,stenull2004commentary,martinoty2004replyT,martinoty2004replyS,menzel2009response}.

For the questions we aim to address here, we will need a different approach. This is because liquid crystal networks, unlike their elastomer counterpart, deform in a non-volume-preserving manner, for which the standard ``neo-classical" approach does not account. Although this can be rectified straightforwardly enough by adding an additional term to the theory, it turns out that the ensuing volume increases are much too small to explain the experimental observations, where volume increases as large as ten percent are reported \cite{liu2017protruding,van2019morphing,van2020electroplasticization,Thesis}.

Another striking difference between liquid crystal networks and liquid crystal elastomers is that the former exhibit a significantly larger density in terms of crosslinks and mesogenic components \cite{white2015programmable}. This suggests that the concept of an anisotropic random walk, on which the ``neo-classical" theory is based, is fundamentally unsuitable for describing liquid crystal networks, as the short spacing between permanent crosslinks does not allow for truly random walks to be explored. This provides a second reason to pursue an alternative approach.

\begin{figure}[htbp]
	\centerline{
		\includegraphics[width=16cm]{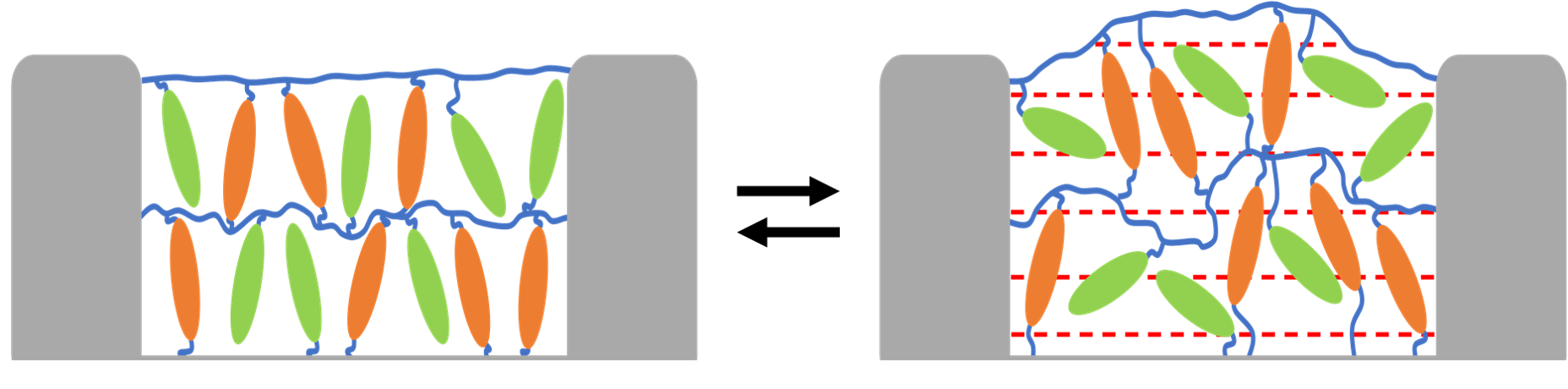}
	}
	\caption{Schematic representation of the liquid crystal network film, sandwiched between two gray electrodes, in the absence (left) and presence (right) of an electric field (dashed red lines). Cross-linked mesogens are indicated in orange, pendant (dipolar) mesogens are indicated in green and polymer strands are indicated in blue. The mesogens are prepared with homeotropic alignment and the black arrows indicate the transition between the ``off" state (left) and the ``on" state (right) of the liquid crystal network, which can be tuned reversibly.}
	\label{fig:diagram geometry}
\end{figure}

Here, we propose a Landau-de Gennes theory for liquid crystal networks, based on the interplay between excluded volume and the volume expansion of the liquid crystal network. Specifically, we focus on electrically-deforming liquid crystal networks, inspired by recent experimental work by Liu and co-workers \cite{liu2017protruding,van2019morphing,van2020electroplasticization}. The corresponding set-up is schematically shown in Figure \ref{fig:diagram geometry}, which indicates that the liquid crystal network comprises a bi-disperse mixture of two distinct mesogen species: (i) relatively immobile mesogens incorporated into the polymer network as permanent crosslinks (orange) and (ii) much more mobile end-on grafted side-group mesogens functionalized with a strong, permanent dipole moment (green). The mesogens are prepared with homeotropic alignment, and upon actuation the electric field (red) is applied in the perpendicular direction, i.e., in the plane. If the electric field is turned on, the material is reported to expand, followed by a viscoelastic relaxation. Upon application of an alternating electric field, steady-state volume expansions as large as ten percent can be achieved.

The focus of this paper is to use this theoretical framework to model and predict the behavior of spatially inhomogeneous systems, with specific application to thin films. The relevant theory to achieve this is developed in Secs. \ref{sec:Theory}-\ref{sec:scalings}. Our point of departure is a stripped-down version of the theory proposed in Ref. \cite{kusters2020dynamical}, featuring a more general interpretation of the mechanism for free volume creation. This results in a more generally-applicable theory, which we illustrate by showing that interpretations based on an electrically-induced glass transition or the collective motion of an array of coupled oscillators could be equally valid. 

Next, in Sec. \ref{sec:phase diagram}, we present the corresponding equilibrium response, which forms the foundation for the relaxational dynamics used in the remainder of this paper. In particular, we establish that the spatial gradients introduced by considering spatially-inhomogeneous systems provide additional pathways for influencing the magnitude of deformation. Following this, in Sec. \ref{sec:permeation}, we demonstrate that the \textit{dynamical} response of the liquid crystal network to a homogeneously-applied, constant electric field starts at the free surface of the film and subsequently permeates the material. We find that the rate at which the response permeates the system is approximately constant in time, and we estimate its dependence on model parameters by means of a short-time approximation, in Sec. \ref{sec:front}; this yields good agreement with our numerical results. Preliminary numerical calculations suggest that upon application of \textit{alternating} electric fields, mimicking those that are used in recent experimental work \cite{liu2017protruding,van2019morphing,van2020electroplasticization}, the same qualitative features persist albeit that they take on more complicated forms. This, in addition to novel features that emerge in response to AC actuation, we intend to analyze in more detail in a future publication \cite{futurework}. In Sec. \ref{sec:interpretations}, we reflect on the different interpretations the theory admits and their relation to experimentally-accessible parameters. We specifically point out the aspect ratio of the mesogens as well as their orientational order when crosslinked into the network as important experimental parameters in the context of permeation, in addition to the electric field strength. Finally, we summarize our most salient findings and remark on their broader significance in Sec. \ref{sec:conclusion}, and we close with an outlook on how the actuation of liquid crystal networks can be further optimized and extended, in Sec. \ref{sec:outlook}.

\section{Model}\label{sec:model system}
The model system we consider, inspired by the experimental work of Liu and co-workers \cite{liu2017protruding,van2019morphing,van2020electroplasticization}, is that of a liquid crystal network dense in terms of mesogens, (almost) all of which are (indirectly) connected to each other by means of short polymer strands. The result is a liquid-crystalline solid. Experimentally, this is achieved by solely using short di-acrylates of the forms =polymer-mesogen-polymer= and =polymer-mesogen, which are added in roughly equal number in the synthesis. Note that this precludes macroscopic phase separation of the mesogenic and polymeric components, and results in two distinct species of mesogen, as schematically illustrated in figure \ref{fig:diagram geometry}. 

One of the mesogen species (orange in the figure) is chemically cross-linked into the polymer network, making it relatively immobile and thus unable to reorient in response to an applied electric field. Conversely, the other species of mesogen (green in the figure) forms end-on grafted pendant side groups that are significantly more mobile. On top of that, this mesogen species is functionalized with a permanent dipole moment, ensuring a strong interaction with an applied electric field. The initial orientational order is such that both mesogen species are homeotropically aligned, whereas the electric field is applied in the perpendicular direction, i.e., in the plane of the film. Finally, the length of the polymer strands connecting the various mesogens in the liquid crystal network is on the order of the mesogen length, and no further permanent crosslinks are present in the polymeric matrix. That is, \textit{all} permanent crosslinks are mesogenic in nature, underscoring the importance of the mesogenic component in this liquid crystal network for both its responsive and mechanical qualities.

The above inspires us to construct a Landau-de Gennes-type theory for this system based on the interplay between excluded-volume generation of the mesogens and the expansion of the liquid crystal network. That is, upon reorientation of the pendant mesogens their mutually-excluded volume with the crosslinked mesogens increases, and with it the total volume of unoccupied space in the liquid crystal network. To also account for the viscoelastic relaxation of this unoccupied volume \cite{liu2017protruding,van2019morphing,van2020electroplasticization}, we add on top of this mechanism a phenomenological relaxation of the liquid crystal network volume. Our approach yields a stripped-down version of the more elaborate theory discussed in Ref. \cite{kusters2020dynamical}, which captures the qualitative dynamics of interest to the current paper.
% This relaxation presumably follows from the polymeric material filling up the pockets of empty space created by the reorientation of the mesogens.

% Here, we present a stripped-down version of this theory, discussed in detail in Appendix B of Ref. \cite{kusters2020dynamical}, as it captures the qualitative dynamics of interest to the current paper.

Although the motivation of the model above is based on excluded-volume arguments, we shall see that this is not explicitly reflected in its final formulation. This suggests we could also arrive at the same---or a similar---theory from a different point of departure. As announced in the introduction, a possible candidate for this is an electrically-induced glass transition, as has recently been suggested by the experimental work of Van der Kooij and co-workers \cite{van2020electroplasticization}. This has in the past already been effectively described in terms of a Landau-type theory similar to ours \cite{gutzow2004glass,gutzow2010glass}.

In the same experimental paper it was hypothesized that collective mesogen reorientation, which potentially weakens the polymer matrix to effect deformation, is at the root of this transition \cite{van2020electroplasticization}. This suggests we could also construct our theory starting with an array of coupled oscillators. In the mean-field description---following the Kuramoto model of coupled oscillators \cite{acebron2005kuramoto}---this approach can be shown to yield a theory of the Ising universality class \cite{daido1994generic,daido1996onset}; we shall see that the model we construct below has the same form. 

% The former has already been effectively described in terms of a Landau-type theory similar to ours in the past \cite{gutzow2004glass,gutzow2010glass}, whereas the critical exponent of the latter --- following the Kuramoto model \cite{acebron2005kuramoto} --- is identical to that of the Ising model \cite{daido1994generic,daido1996onset}.

% This suggests the order parameter $f$ could in those cases instead represent a measure for the difference in molar entropy between the glassy and the rubbery state, or the fraction of oscillators that oscillate collectively, respectively. To keep the presentation throughout this paper as generic as possible we shall not hone in on one specific interpretation, but only discuss the different interpretations and their connections with experimentally-accessible parameters in more detail in Sec. \ref{sec:conclusion}, after showing our generally-applicable results.

The following sections are devoted to developing the theory that we shall investigate for the remainder of this paper. Readers solely interested in the results can skip to Sec. \ref{sec:phase diagram}, where we discuss the equilibrium response of the model. We shall return to the different model interpretations and their connections with experimentally-accessible parameters further below in Sec. \ref{sec:interpretations}, after showing our generally-applicable results.

% To keep the presentation throughout this paper as generic as possible we shall not hone in on one specific interpretation, but only discuss the different interpretations and their connections with experimentally-accessible parameters in more detail in Sec. \ref{sec:conclusion}, after showing our generally-applicable results.

\section{Bulk theory}\label{sec:Theory}

The theory describes the state of the liquid crystal network in terms of two key order parameters. For now, we neglect the viscoelastic relaxation of the polymer network; we return to this below when we discuss the model dynamics. The first of the order parameters quantifies the orientation of the different mesogen species. Since only one of these is able to reorient in response to an applied electric field---the other being fully crosslinked into the polymer network---we focus on the pendant (dipolar) mesogens. Our Molecular Dynamics (MD) computer simulations have indicated that upon application of the electric field these mesogens may either (i) reorient to align with the electric field, perpendicular to the crosslinked mesogens, or (ii) be impeded in their reorientation by their crosslinked neighbors, thus remaining along their initial axis of orientation \cite{kusters2020dynamical}. We capture this mechanism by introducing a two-population model in terms of the order parameter $f$, where $0\leq f^2\leq1$ indicates the fraction of pendant mesogens that aligns with the electric field. The order parameter $f$ enters only in even powers to permit its consistent interpretation as a (positive) fraction.

The model reflects a competition between the electric field, with field strength $H$ \footnote{The form $H\propto\lvert\underline{E}\rvert^2$ is appropriate for describing the free energy of an induced dipole due to the inversion symmetry of the nematic director $\underline{n}\rightarrow-\underline{n}$. Although for a mesogen with permanent dipole moment $\underline{p}$ the analogous expression would read $-\underline{p}\cdot\underline{E}$, averaging this over an ensemble of mesogens using a Boltzmann distribution this can again be written in the form $-\langle\underline{p}\cdot\underline{E}\rangle\propto\lvert\underline{E}\rvert^2$ for sufficiently weak electric fields \cite{prost1995physics,de2012liquid,vertogen2012thermotropic}}, favoring reorientation of the pendant mesogens on the one hand, and the excluded-volume interactions between the different mesogen species opposing this reorientation on the other. This suggests a critical field strength, $H_*$, that must be overcome in order to make reorientation energetically favorable. As a matter of fact, the existence of a critical field emerges naturally from the more elaborate theory mentioned earlier \cite{kusters2020dynamical}. Its value depends on the aspect ratio of the mesogens, their orientational order at the time of crosslinking and the crosslinking fraction of the network. The associated contribution to the Gibbs free energy per unit \textit{reference} volume then reads \cite{kusters2020dynamical} \footnote{If we were to write down the \textit{actual} Gibbs free energy density $\mathcal{G}=G/V$, with $G$ the total Gibbs free energy, thermodynamic equilibrium would demand $\partial \mathcal{G}/\partial\eta=-\mathcal{G}/\left(1+\eta\right)$. This means the Gibbs free energy density must take the form $\mathcal{G}=g/\left(1+\eta\right)$, with $\partial g/\partial \eta=0$. This indicates that, if we wish to treat $\eta$ as a proper order parameter, the Gibbs free energy per unit \textit{reference} volume, $g= G/V_0$, is the relevant thermodynamic potential. We explicitly \textit{construct} the Landau theory to yield the proper behavior upon minimization with respect to all order parameters}
\begin{equation}\label{eq:g1}
    g_1=\frac{1}{2}\left(H_*-H\right) f^2+\frac{1}{4}B_ff^4,
\end{equation}
where we have also added a bulk-modulus-like term, with phenomenological parameter $B_f$. This term penalizes field-orientation of the pendant mesogens, as this induces local matrix strain. Although such a term usually serves to ensure the free energy is bounded from below, here this criterion is already satisfied by the constraint $0\leq f^2\leq 1$. Instead, we require this term to allow $f^2$ to take intermediate values between $0$ and $1$ upon minimization of Eq. \eqref{eq:g1} with respect to $f$.

Subsequently, we couple Eq. \eqref{eq:g1} to the scaled volume expansion of the liquid crystal network, $\eta\equiv\left(V-V_0\right)/V_0$, with $V$ the volume of the liquid crystal network and $V_0$ its initial volume, by relating the degree of orientational order of the mesogens to the volume of the liquid crystal network. More specifically, we associate global orientational disorder of the mesogens, e.g., with the pendant mesogens oriented perpendicularly with respect to the cross-linked mesogens, with a large volume, whereas we associate global orientational order of the mesogens, e.g., with the pendant mesogens aligned with the cross-linked mesogens, with a small volume. Using $\xi$ as the phenomenological coupling constant, we then write the associated contribution to the Gibbs free energy per unit \textit{reference} volume as
\begin{equation}\label{eq:g2}
    g_2=-\xi\eta f^2+\frac{1}{2}B_\eta \eta^2.
\end{equation}
Here, we have again added a bulk-modulus-like term, with phenomenological coefficient $B_\eta$, which guarantees that the free energy remains bounded from below. The bulk moduli $B_f$ and $B_\eta$, together with the coupling constant $\xi$, dictate the effective resistance of the system to deformations, which can be made explicit by integrating out the volume-expansion order parameter $\eta$.

The total free energy per unit \textit{reference} volume is then $g=g_1+g_2$, where we have written no explicit pressure-volume contribution as the pressure can effectively be absorbed in our phenomenological parameters (not shown). The free energy is reminiscent of the Ising model and can be mapped exactly onto this class of model by integrating out the volume-expansion order parameter. This, however, only applies in thermodynamic equilibrium and neglects the geometric significance of $\eta$, hence the choice to treat both order parameters explicitly in this paper.

This concludes the development of the bulk theory, which we extend to a spatially-inhomogeneous description below.

\section{Inhomogeneous theory}\label{sec:Theory2}

To pave the way for an investigation of the permeation properties of liquid crystal network coatings, we now extend the local theory by considering a thin film in one spatial dimension, that is, in a direction perpendicular to the plane of the thin film. Prior to the application of an electric field, the film extends from $z_0=0$ to $z_0=L_0$, which describes the undeformed, \textit{reference} configuration. In this (fictitious) configuration the liquid crystal network is presumed to be prepared homogeneously, such that it contains no spatial gradients. The subscript ``$0$" serves to distinguish this from the deformed, \textit{target} configuration that is obtained upon applying an electric field. We pay particular attention to the mapping between these configurations because our order parameters carry geometric information, meaning that this mapping can act as a proxy for additional order-parameter couplings.

To see this, consider that the volume-expansion order parameter, $\eta\left(z_0\right)$, represents the relative expansion of a volume element located at height $z_0$ in the reference configuration. In the target configuration, this volume element is translated to a greater height due to the expansion of the volume elements located below it, such that we recover the height co-ordinate in the target configuration as
\begin{equation}\label{eq:conformalmap}
    z\left(z_0\right)=\int_0^{z_0}\diff\xi\left(1+\eta\left(\xi\right)\right).
\end{equation}
In writing Eq. \eqref{eq:conformalmap} we have implicitly assumed \textit{all} volume increases add to the thickness of the thin film, rather than spreading it out laterally. This assumption is justified because the experimental system we aim to model is clamped in the lateral plane, and so can only expand in the direction of the film height \cite{liu2017protruding,van2019morphing,van2020electroplasticization}. Although we might then expect wrinkling instabilities to play an important role, it can be shown that these only become significant for volume changes far exceeding those we consider here \cite{tanaka1987mechanical,onuki1989theory,trujillo2008creasing}. Although the inverse mapping, $z_0\left(z\right)$, also exists, below we show that we do not require its explicit form and so need not dwell on it. 

By virtue of Eq. \eqref{eq:conformalmap}, we now write the total Gibbs free-energy functional per unit area $A$ as an integral over the thickness of the film, according to \footnote{This expression can be derived by dividing the (reference) system into elements and ascribing each a free energy. If we subsequently add a free-energy penalty for neighboring elements that have deviating order parameter values, the sum over all elements can be written in the given form}
\begin{equation}\label{eq:Gibbstarget} 
    \frac{G}{A}=\int_0^{L}\diff z \, \frac{1}{1+\eta\left(z_0\left(z\right)\right)}\Bigg[g+\frac{\kappa_f^2}{2}\left(\frac{\partial f\left(z_0\left(z\right)\right)}{\partial z}\right)^2
    +\frac{\kappa_\eta^2}{2}\left(\frac{\partial\eta\left(z_0\left(z\right)\right)}{\partial z}\right)^2\Bigg].
\end{equation}
Here, we have tacitly added square-gradient contributions, with phenomenological coefficients $\kappa_f$ and $\kappa_\eta$, to ensure smooth order parameter profiles, and we require the prefactor $1/\left(1+\eta\left(z_0\left(z\right)\right)\right)$ to conserve the density integral of our system. Note that, generally, the square gradient coefficients $\kappa_f$ and $\kappa_{\eta}$ are not independent of the other model parameters as they result from local interactions within the theory \cite{de1979scaling}. Nevertheless, we shall for simplicity treat them as free parameters until we discuss their experimental interpretation at the end of this paper.

% \textbf{We associate the square gradient coefficients with experimentally-accessible parameters by noting that they result from local interactions within our theory. This suggests that $K_f$ increases with the mesogen dimensions, $l^2d$, as well as the as-prepared degree of orientational order of the mesogens, whereas $K_\eta$ scales with the cross-linking fraction of the polymer network.}

Although Eq. \eqref{eq:Gibbstarget} looks deceptively compact, many of its complexities are ``hidden" in the mapping. For example, the film thickness, $L$, follows directly from Eq. \eqref{eq:conformalmap}, making explicit the moving-boundary problem. Similarly, all differentials of $z$ carry a dependence on the local volume expansion. Finally, all order parameters, which the local theory defines relative to the reference configuration, are given in terms of the inverse mapping $z_0\left(z\right)$, reflecting that all volume elements must be translated appropriately upon transformation to the target configuration. We address these complexities by mapping the problem onto the reference configuration, which transforms Eq. \eqref{eq:Gibbstarget} to
\begin{equation}\label{eq:Gibbsreference}
    \frac{G}{A}=\int_0^{L_0}\diff z_0 \Bigg[g+\frac{\kappa_f^2}{2\left(1+\eta\left(z_0\right)\right)^2}\left(\frac{\partial f\left(z_0\right)}{\partial z_0}\right)^2
    +\frac{\kappa_\eta^2}{2\left(1+\eta\left(z_0\right)\right)^2}\left(\frac{\partial\eta\left(z_0\right)}{\partial z_0}\right)^2\Bigg],
\end{equation}
where all order-parameter dependencies are made explicit.

Subsequently, we recover the equilibrium configuration of the liquid crystal network film at a given electric field strength by functionally minimizing Eq. \eqref{eq:Gibbsreference} with respect to both order parameters, subject to the following boundary conditions. At the bottom of the film, at $z_0=0$, the liquid crystal network is clamped to a substrate, and thus unable to expand. As a result, we impose the Dirichlet boundary conditions $f\left(0\right)=\eta\left(0\right)=0$.

Conversely, at the top of the film, at $z_0=L_0$, the liquid crystal network is free to expand and has an interface with the ambient medium. We assume that across this interface the density of the liquid crystal network varies continuously. Consequently, we demand that the volume-expansion order parameter---which represents an inverse density---assumes its maximum value at the upper boundary of the film; this effectively models a diffuse interface. Based on the phase diagram of the bulk theory \cite{kusters2020dynamical}, this leads us to simultaneously demand $\eta\left(L_0\right)=\xi/B_\eta$ and $f\left(L_0\right)=1$. Although other choices of boundary condition are possible for $f$ at this interface, these do not significantly alter the permeation phenomenon we highlight in this paper. We thus choose our boundary conditions under the aspect of simplicity, and the transparency of the results.

This completes the description of our spatially-inhomogeneous extension to the theory. Below, in Sec. \ref{sec:dynamics}, we build upon this framework to allow dynamics, and thus facilitate studying the time-resolved behavior of the liquid crystal network.

\section{Relaxational dynamics}\label{sec:dynamics}

To extend the equilibrium theory we discuss in the previous sections to dynamics, we introduce a set of differential equations that dictate the evolution of the order parameter profiles as a function of time, $t$. The simplest form we can choose for this describes relaxational dynamics, according to \cite{hohenberg1977theory}
\begin{equation}\label{eq: dynamics}
\begin{split}
    \partial_{t}{f}&=-{\Gamma}_{f}\frac{\delta {G}}{\delta {f}},\\
    \partial_{t}{\eta}&=-{\Gamma}_{{\eta}}\frac{\delta {G}}{\delta {\eta}}-{\gamma} \, \left({\eta}-\eta_0\right) \, \tau\left(z_0,t\right).
\end{split}
\end{equation}
Here, $\Gamma_f$ and $\Gamma_{\eta}$ represent kinetic coefficients, which encapsulate the dissipative processes by which the system can decrease its free energy upon varying the relevant order parameter. In addition to the mean-field dynamics that follows from the free-energy-functional Eq. \eqref{eq:Gibbsreference}, on the second line we have also imposed a term reflecting the viscoelastic relaxation of the polymer network. This phenomenological term, with coefficient $\gamma$, ensures that the volume-expansion order parameter relaxes back to its initial profile in the absence of an electric field, $\eta\left(z_0\right)=\eta_0\left(z_0\right)$, as a function of time. This profile generally differs from the profile that minimizes the free-energy-functional Eqn. \eqref{eq:Gibbsreference} in the \textit{presence of an electric field}; this choice of enforced equilibrium is informed by experiments and Molecular Dynamics computer simulations \cite{liu2017protruding,van2019morphing,van2020electroplasticization,kusters2020dynamical}. 

Although this relaxation itself can be achieved by simply multiplying the term with the time $t$---such a term invariably dominates the dynamics in the long-time limit---this approach breaks down in the case of alternating electric fields. Then, the response to each subsequent field oscillation would be increasingly suppressed, meaning a dynamic steady state cannot exist.
To address this, we introduce the relaxation function $\tau\left(z_0,t\right)$, which is a highly nontrivial contribution to the theory that acts as a memory function of the system. The key insight here, is that to properly describe the relaxation of a volume element $\eta\left(z_0,t\right)$, we must keep track of both the time since it was perturbed from its initial value and the additional perturbations that have been applied since then.
The relaxation function $\tau\left(z_0,t\right)$ combines these quantities to yield a measure of how strongly a given volume element should relax at a given time, which we make numerically accessible by recursively defining
\begin{equation*}
    \tau\left(t,z_0\right)=\frac{\eta\left(z_0,t-\Delta t\right)\left[t- \tau\left(z_0,t-\Delta t\right)\right]+\lvert\eta\left(z_0,t\right)-\eta\left(z_0,t-\Delta t\right)\rvert \, t}{\eta\left(z_0,t\right)},
\end{equation*}
with $\Delta t$ the numerical time increment. This last term provides an effective dynamical coupling between the order parameters that goes beyond the free-energetic nature of the Landau theory.

In writing Eqn. \eqref{eq: dynamics} we have omitted a Gaussian noise term, which is normally added to ensure the fluctuation-dissipation theorem is satisfied \cite{glauber1963time}. Although such a term is known to play an important role in various systems \cite{hamada1981dynamics,van1994noise,park1996noise,van1997nonequilibrium,lee2014critical}, we justify focusing solely on the deterministic part of the dynamics by pointing out that the phase diagram we reported in Ref. \cite{kusters2020dynamical} does not exhibit any free-energy barriers; we shall again encounter this when we discuss our equilibrium response, in Sec. \ref{sec:phase diagram}. This suggests there are no kinetic traps, meaning the system will invariably reach the proper equilibrium configuration. Our choice of boundary conditions ensures that the initial equilibrium configuration can, upon application of an electric field, be left without needing to incorporate noise.

The above represents the final extension of the theory we discuss in this paper. However, before we analyze the results, below we first reduce the parameter space of the model by scaling the theory to make it dimensionless. 

% the absence of any free-energy barriers in the phase diagram we report in figure \ref{fig:phase diagram}, which we verified by performing extensive numerical sweeps \cite{kusters2020dynamical}, suggests this is not the case for the current system. That is, in the absence of such barriers there are no kinetic traps, such that the system invariably ends up in the proper equilibrium configuration. This justifies us dropping the Gaussian noise term from Eq. \eqref{eq: dynamics}, and focusing solely on its deterministic part.

% We remark that although the Gaussian noise term, $\theta_\psi$, is known to play an important role in various systems \cite{hamada1981dynamics,van1994noise,park1996noise,van1997nonequilibrium,lee2014critical}, the absence of any free-energy barriers in the phase diagram we report in figure \ref{fig:phase diagram}, which we verified by performing extensive numerical sweeps \cite{kusters2020dynamical}, suggests this is not the case for the current system. That is, in the absence of such barriers there are no kinetic traps, such that the system invariably ends up in the proper equilibrium configuration. This justifies us dropping the Gaussian noise term from Eq. \eqref{eq: dynamics}, and focusing solely on its deterministic part.

\section{Scaling procedure}\label{sec:scalings}
To scale the theory, we introduce the scaled volume-expansion order parameter $\tilde{\eta}\equiv\eta/\eta\left(L_0\right)$, whereas we leave the population order parameter $f$ unchanged to ensure we can explicitly demand $0\leq f^2\leq1$. Subsequently treating the critical field strength $H_*$ as the characteristic energy scale of the system, we identify the scaled field strength $h= \left(H-H_*\right)/H_*$, as well as the scaled bulk modulus $\tilde{B}_f= B_f/H_*$ and coupling constant $\zeta=\xi^2/B_\eta H_*$. Upon also scaling the spatial coordinate to the reference thickness of the film, $\tilde{z}_0=z_0/L_0$, the scaling procedure for the equilibrium theory is completed by recognizing the dimensionless square gradient coefficients $\tilde{\kappa}_f= \kappa_f/\sqrt{H_*L_0^2}$ and $\tilde{\kappa}_{\tilde{\eta}} =   \kappa_\eta/\sqrt{H_*B_{\eta}^2L_0^2/\xi^2}$.

% Following this, we define the scaled field strength $h\equiv \left(H-H_*\right)/H_*$, as well as identify the scaled phenomenological parameters $\tilde{B}_f\equiv B_f/H_*$ and $\zeta\equiv\xi^2/B_\eta H_*$. Upon also scaling the spatial co-ordinate $\tilde{z}_0=z_0/L_0$, the scaling procedure for the (non-local) equilibrium theory is completed by recognizing the dimensionless square gradient coefficients $\tilde{\kappa}_f\equiv \kappa_f/\sqrt{H_*L_0^2}$ and $\tilde{\kappa}_{\tilde{\eta}} \equiv   \kappa_\eta/\sqrt{H_*B_{\eta}^2L_0^2/\xi^2}$.

To identify the relevant dynamical parameters, next we remark that the time scale of interest to us is that on which the liquid crystal network deforms. To make this explicit, we scale the time such that it is measured relative to the dynamics of the volume-expansion order parameter $\tilde{\eta}$, i.e., we demand that the corresponding effective kinetic coefficient reads $\tilde{\Gamma}_{\tilde{\eta}}=1$. From this choice it follows that a natural scaling for the time is $\tilde{t}=tB_{\eta}^2H_*\Gamma_{\eta}/\xi^2$; similarly, the scaled relaxation function becomes $\tilde{\tau}=\tau B_{\eta}^2H_*\Gamma_{\eta}/\xi^2$. The effective kinetic coefficient of the population order parameter, $\tilde{\Gamma}_f=\Gamma_f\xi^2/\Gamma_{\eta}B_{\eta}^2$, and the effective coefficient for viscoelastic relaxation, $\tilde{\gamma}=\gamma\xi^4/B_{\eta}^4H_*^2\Gamma_{\eta}^2$, can then be straightforwardly read off.
We choose our parameter values in the broad regime $\tilde{\Gamma}_f<\tilde{\Gamma}_{\tilde{\eta}}$ that we have previously shown to be qualitatively consistent with computer simulations and experiments \cite{kusters2020dynamical}.

% To identify the relevant dynamical parameters, next we non-dimensionalize Eq. \eqref{eq: dynamics} by scaling the time $\tilde{t}=tB_{\eta}^2H_*\Gamma_{\eta}/\xi^2$. It is then straightforward to read off the effective kinetic coefficients $\tilde{\Gamma}_{\tilde{\eta}}=1$ and $\tilde{\Gamma}_f=\Gamma_f\xi^2/\Gamma_{\eta}B_{\eta}^2$, as well as the effective coefficient for viscoelastic relaxation $\tilde{\gamma}=\gamma\xi^4/B_{\eta}^4H_*^2\Gamma_{\eta}^2$. This shows that our scaled theory specifically measures time relative to the dynamics of the volume-expansion order parameter $\tilde{\eta}$, and we choose our parameter values in the broad regime $\tilde{\Gamma}_f<\tilde{\Gamma}_{\tilde{\eta}}$ that we have previously shown to be qualitatively consistent with computer simulations and experiments \cite{kusters2020dynamical}.

This finalizes the discussion on our theory. We now present the corresponding equilibrium response, which forms the foundation for the relaxational dynamics we require to study the time evolution of the non-local volume expansion.

\section{Equilibrium response}\label{sec:phase diagram}
\begin{figure}[htbp]
    \subfloat[]{\includegraphics[width=8.cm]{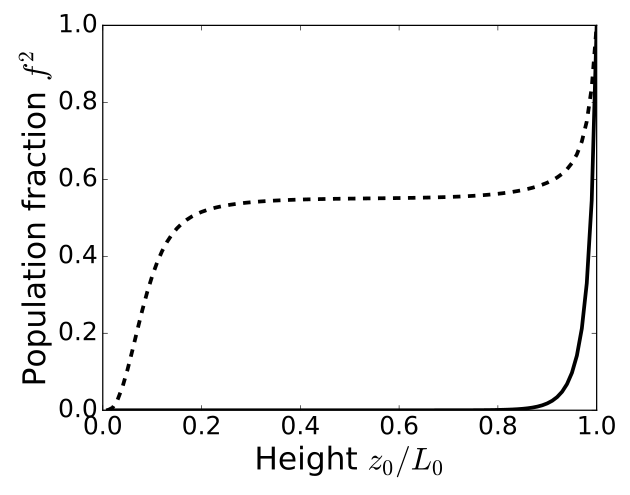}}
    \subfloat[]{\includegraphics[width=8.cm]{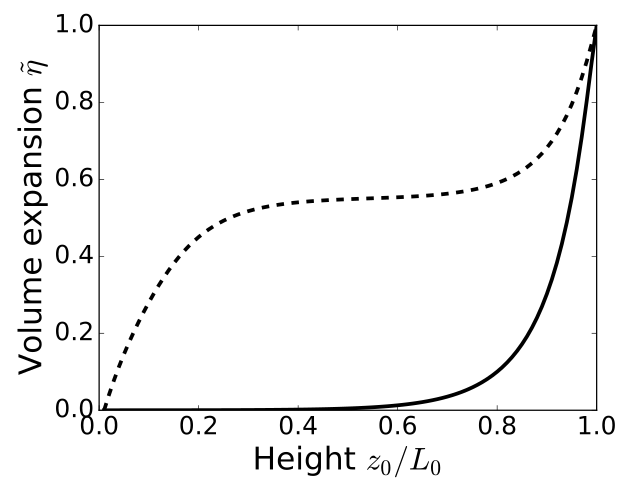}} \\
    \caption{Fraction of mesogens that respond collectively to an applied electric field (a) and the relative volume increase of the liquid crystal network (b), obtained from the scaled theory and as a function of the scaled height co-ordinate ${z}_0/L_0$. The solid curves indicate the order parameter profiles in the absence of an electric field, whereas the dashed curves correspond to a super-critical scaled field strength $h=1.48$. We denote the population fraction order parameter $f^2$, and the scaled volume-expansion order parameter $\tilde{\eta}$. See Sec. \ref{sec:scalings} for the scaling procedure. The values used for the scaled model parameters are $\zeta=0.51$ for the mesogen-volume coupling constant, $\tilde{B_f}=3.70$ for the bulk-modulus-like free-energy penalty for the population fraction $f^2$, $\eta\left(L_0\right)=0.675$ for the reference value used to scale the volume-expansion order parameter ${\eta}$ (obtained through a fit to experiments, see the main text for details), and $\tilde{\kappa}_f=0.071$ and $\tilde{\kappa}_{\tilde{\eta}}=0.071$ for the square-gradient coefficients of the corresponding order parameters.}\label{fig:illustrate}
\end{figure}

We commence the analysis of our theory by investigating the equilibrium profiles our order parameters assume within the thin-film geometry. To this end, figure \ref{fig:illustrate} shows the results of a numerical minimization of the free-energy-functional Eqn. \eqref{eq:Gibbsreference}, scaled as described in Sec. \ref{sec:scalings} and subject to the boundary conditions $f\left(0\right)=\tilde{\eta}\left(0\right)=0$, $f\left(1\right)=\tilde{\eta}\left(1\right)=1$; the chosen parameter values correspond to a broad range that qualitatively reproduces key features of both MD computer simulations and experiments \cite{liu2017protruding,kusters2020dynamical}. In following the above procedure we neglect, for now, the viscoelastic relaxation of the polymer network, which would otherwise destroy any volume increase at thermodynamic equilibrium; we do incorporate this below when we discuss dynamics.

The order parameter profiles shown in figure \ref{fig:illustrate} physically represent the fraction of mesogens that respond collectively to the electric field (e.g., by aligning with it or reorienting coherently), $0\leq f^2\leq1$ (left), and the corresponding volume expansion of the liquid crystal network $\tilde{\eta}$ (right), where the latter is scaled specifically such that $\tilde{\eta}\left(L_0\right)=1$ holds. The solid curves correspond to the profiles in the absence of an electric field ($h=-1$), whereas the dashed curves result from the application of a super-critical electric field ($h>0$). Recall that here the electric field strength $h$ measures the scaled distance to the critical field value, i.e., the value at which it locally becomes free-energetically favorable for the mesogens to respond to the field.

From the solid curves in figure \ref{fig:illustrate} we conclude that in the absence of an electric field both order parameter profiles are characterized by (i) a broad region of no response (there is no stimulus to respond to), and (ii) a thin boundary layer near the surface of the film, where spatial gradients become important by virtue of the boundary conditions $f\left(1\right)=\tilde{\eta}\left(1\right)=1$. If we subsequently apply a super-critical electric field (dashed curves), a bulk region emerges in the center of the film, dominated by the local response of the liquid crystal network. Consequently, a second boundary layer is necessary at the bottom of the film to ensure the boundary conditions imposed by the substrate, $f\left(0\right)=\tilde{\eta}\left(0\right)=0$, are satisfied. Note that even if we were to choose different, e.g., Neumann boundary conditions at the surface of the film, this general structure of two boundary layers flanking the bulk region persists. The findings in this paper we find to be robust provided that $f\left(1\right)=\tilde{\eta}\left(1\right)>0$, i.e., the symmetry between the top and bottom of the film remains broken (results not shown).

\begin{figure}[htbp]
    {\includegraphics[width=8.cm]{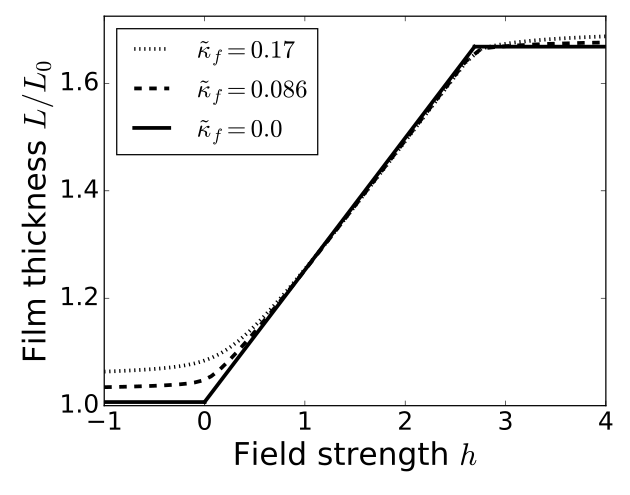}}\\
    \caption{Equilibrium thickness of the liquid crystal network film obtained from the scaled theory, as a function of the scaled field strength $h$. Different curves correspond to different values of the scaled square gradient coefficient $\tilde{\kappa}_f$, while $\tilde{\kappa}_{\tilde{\eta}}$ is set to zero. See Sec. \ref{sec:scalings} for the scaling procedure and see figure \ref{fig:illustrate} for the parameter values not explicitly stated here.}\label{fig:phase diagram f}
\end{figure}

Next, we investigate how the order parameter profiles change upon varying the field strength $h$ and the square gradient coefficients $\tilde{\kappa}_f$ and $\tilde{\kappa}_{\tilde{\eta}}$. The former dictates the local response of the liquid crystal network and so favors a broad bulk-like response with very narrow boundary layers to impose the boundary conditions. Conversely, the latter penalize spatial gradients in their respective order parameters and so generally favor smoothly-varying profiles with extended boundary layers. (For a brief discussion of the associated length scales, see Appendix \ref{app:lengthscales}.) We focus specifically on their influence on the expansion of the film thickness $L/L_0=\int_0^1\diff\tilde{z}_0 \left(1+\eta\left(L_0\right) \, \tilde{\eta}\right)$, which provides a general overview of our results, where the parameter $\eta\left(L_0\right)=0.675$ is determined by means of a fit to experiments \cite{liu2017protruding}. Figure \ref{fig:phase diagram f} shows the results as a function of $h$, where the different curves correspond to different values of $\tilde{\kappa}_f$; we discuss the effect of varying $\tilde{\kappa}_{\tilde{\eta}}$ further below, which is set to zero here.

The solid curve in figure \ref{fig:phase diagram f} indicates the local limit of our theory, i.e., $\tilde{\kappa}_f=\tilde{\kappa}_{\tilde{\eta}}=0$. From this we recognize the critical field strength at $h=0$, below which there is no response to the electric field. Then, as we increase the electric field strength beyond its critical value, the film thickness grows linearly with $h$, until it saturates around $h\approx2.7$. The linear growth of the film thickness with the field strength follows directly from the fact that the population fraction order parameter, $f^2$, and in turn also the volume-expansion order parameter, $\tilde{\eta}$, responds directly to $h$, as can be seen from Eq. \eqref{eq:g1}. Similarly, the saturation of the film thickness follows from the fact that $f^2$ represents a population \textit{fraction}, which we constrain to $0\leq f^2\leq1$; this consequently bounds the volume expansion as well. Note that this also explains the sharpness of the transitions apparent from figure \ref{fig:phase diagram f}: at $h=0$ a continuous phase transition occurs, whereas at $h\approx2.7$ we invoke the physical upper bound $0\leq f^2\leq1$. Finally, the reason the saturated thickness of the film, $L/L_0=1.675$, far exceeds the deformations of up to ten percent reported in experiments is that we have yet to incorporate the viscoelastic relaxation of the polymer network \cite{liu2017protruding} \footnote{Viscoelastic relaxation implies that no relative volume increases can persist in thermodynamic equilibrium; in experiments the liquid crystal is kept perpetually out of equilibrium by means of alternating electric fields. The fit takes this into account}.

The first deviations from the local theory can be seen from the dashed and dotted curves in figure \ref{fig:phase diagram f}, which correspond to progressively larger values of $\tilde{\kappa}_f$. These show that for $\tilde{\kappa}_f>0$ the film thickness already exceeds the reference thickness \textit{in the absence of an electric field}, i.e., $L/L_0>1$ for $h=-1$, indicating that the initial configuration is already deformed relative to the (fictitious) homogeneous reference configuration defined in Sec. \ref{sec:Theory2}. This is because $\tilde{\kappa}_f$ penalizes spatial gradients in the population fraction $f^2$, leading to an increasingly smooth order parameter profile; the same holds for the volume-expansion order parameter $\tilde{\eta}$, since the two are coupled. This implies a widening of the boundary layer near the surface of the film, where $f^2,\tilde{\eta}>0$ (see figure \ref{fig:illustrate}); the result is an increase in the thickness integral $L/L_0=\int_0^1\diff\tilde{z}_0 \left(1+\eta\left(L_0\right) \, \tilde{\eta}\right)$. That is, by virtue of the gradient contributions to the theory, in the absence of an electric field the film thickness, $L$, exceeds the thickness of a perfectly homogeneous reference film, $L_0$.

In addition, figure \ref{fig:phase diagram f} shows that the critical field strength we recognized from the solid curve becomes increasingly smooth as the value of $\tilde{\kappa}_f$ is increased (from the solid to the dashed to the dotted curve), allowing expansion far below $h=0$. This is because $h=0$ is only a critical field in the local sense, representing the required electric field strength to induce reorientation and concomitant expansion away from the configuration $f^2=\tilde{\eta}=0$. The \textit{effective} electric field strength depends on the local value of the order parameters $f^2$ and $\tilde{\eta}$, and thus is influenced by the increasingly smooth order parameter profiles discussed above; this can be seen directly from Eqs. \ref{eq:g1} and \ref{eq:g2}, where the total coefficient of the $f^2$ term contains both the electric field $H$ and the relative volume expansion $\eta$. The $f$-dependence stems from higher-order terms in the Landau free energy.

Finally, we find that the saturation thickness of the film is an increasing function of $\tilde{\kappa}_f$. This we rationalize by noting that spatial gradients in the population fraction $f^2$ can be ``smeared out" by locally expanding the liquid crystal network; this can also be seen from the prefactor $1/\left(1+\eta\right)^2$ in Eqn. \eqref{eq:Gibbsreference}.

% Finally, we find that the saturation thickness of the film is an increasing function of $\tilde{\kappa}_f$. We rationalize this by noting that spatial gradients in the population fraction $f^2$ can be ``smeared out" by locally expanding the liquid crystal network; this can also be seen from the prefactor $1/\left(1+\eta\right)^2$ in Eq. \eqref{eq:Gibbsreference}. Note that since figure \ref{fig:phase diagram f} shows a film thickness $L/L_0>1.675$ for the dashed and dotted curves, we conclude that this ``gradient coupling" between the order parameters induces --- at least locally --- an expansion that exceeds the saturated value of the bulk theory. \textbf{Vaag, en bovendien nietszeggend als je de theorie niet leest...}

\begin{figure}[htbp]
    \subfloat[]{\includegraphics[width=8.cm]{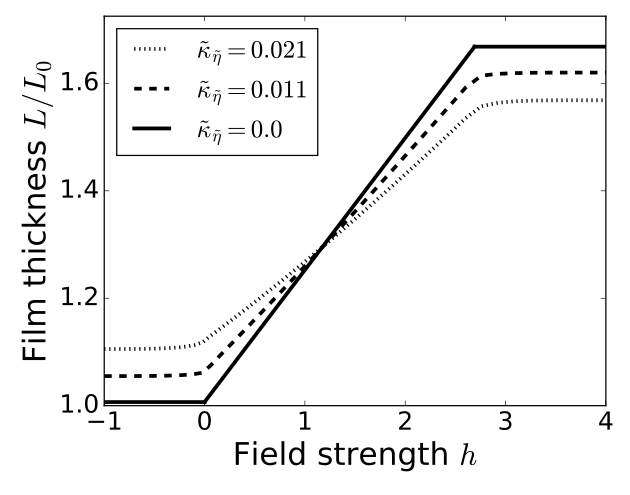}}\\
    \caption{Equilibrium thickness of the liquid crystal network film obtained from the scaled theory, as a function of the scaled field strength $h$. Different curves correspond to different values of the scaled square gradient coefficient $\tilde{\kappa}_{\tilde{\eta}}$, while $\tilde{\kappa}_f$ is set to zero. See Sec. \ref{sec:scalings} for the scaling procedure and see figure \ref{fig:illustrate} for the parameter values not explicitly stated here.}\label{fig:phase diagram eta}
\end{figure}

Many of the qualitative features we discuss above also hold upon varying $\tilde{\kappa}_{\tilde{\eta}}$ at fixed $\tilde{\kappa}_f=0$, as figure \ref{fig:phase diagram eta} illustrates. Indeed, the solid curve again corresponds to the local limit $\tilde{\kappa}_f=\tilde{\kappa}_{\tilde{\eta}}=0$, and upon increasing $\tilde{\kappa}_{\tilde{\eta}}$ (from the solid to the dashed to the dotted curve) the film thickness at zero electric field strength increases. However, we find that the saturation thickness of the film \textit{decreases}, rather than increases, upon increasing $\tilde{\kappa}_{\tilde{\eta}}$. This we can trace back to the electric field not acting directly on the volume of the liquid crystal network, but only indirectly via the population fraction $0\leq f^2\leq1$. As a result, the driving force behind volume expansion remains constrained, which, if we also recall that increasing $\tilde{\kappa}_{\tilde{\eta}}$ makes it increasingly costly to locally induce volume increases, yields the decreased expansion shown in figure \ref{fig:phase diagram eta}.

This completes the discussion on our equilibrium response, which shows that, in addition to the electric field strength, the square gradient coefficients also provide pathways to influence the (local) deformation of the liquid crystal network. In the following section, we investigate the relaxational dynamics that follow from the equilibrium theory discussed above, with a particular focus on the permeation of the response into the thin film. Here, we explicitly incorporate the viscoelastic relaxation of the polymer network, as detailed in Sec. \ref{sec:dynamics}.

% This completes the discussion on our equations of state, which provide qualitative insight into the mechanisms underlying liquid crystal network expansion. In particular, we find that, in addition to the electric field strength, the square gradient coefficients also provide pathways to influence the extent to which the film expands; these pathways distinguish the non-local theory from its local counterpart. 

% \textbf{The major omission in this narrative, however, is the viscoelastic relaxation of the polymer network, which fills up the empty space freed up by the mesogens as a function of time. We address this below in Sec. \ref{sec:permeation}, which is dedicated to the dynamical behavior of the model. This framework we use to study the permeation of the response into the liquid crystal network, resolved in both space and time.
% }

% \section{Dynamical response}\label{sec:permeation}
\section{Permeation}\label{sec:permeation}
\begin{figure}[htbp]
    {\includegraphics[width=8.cm]{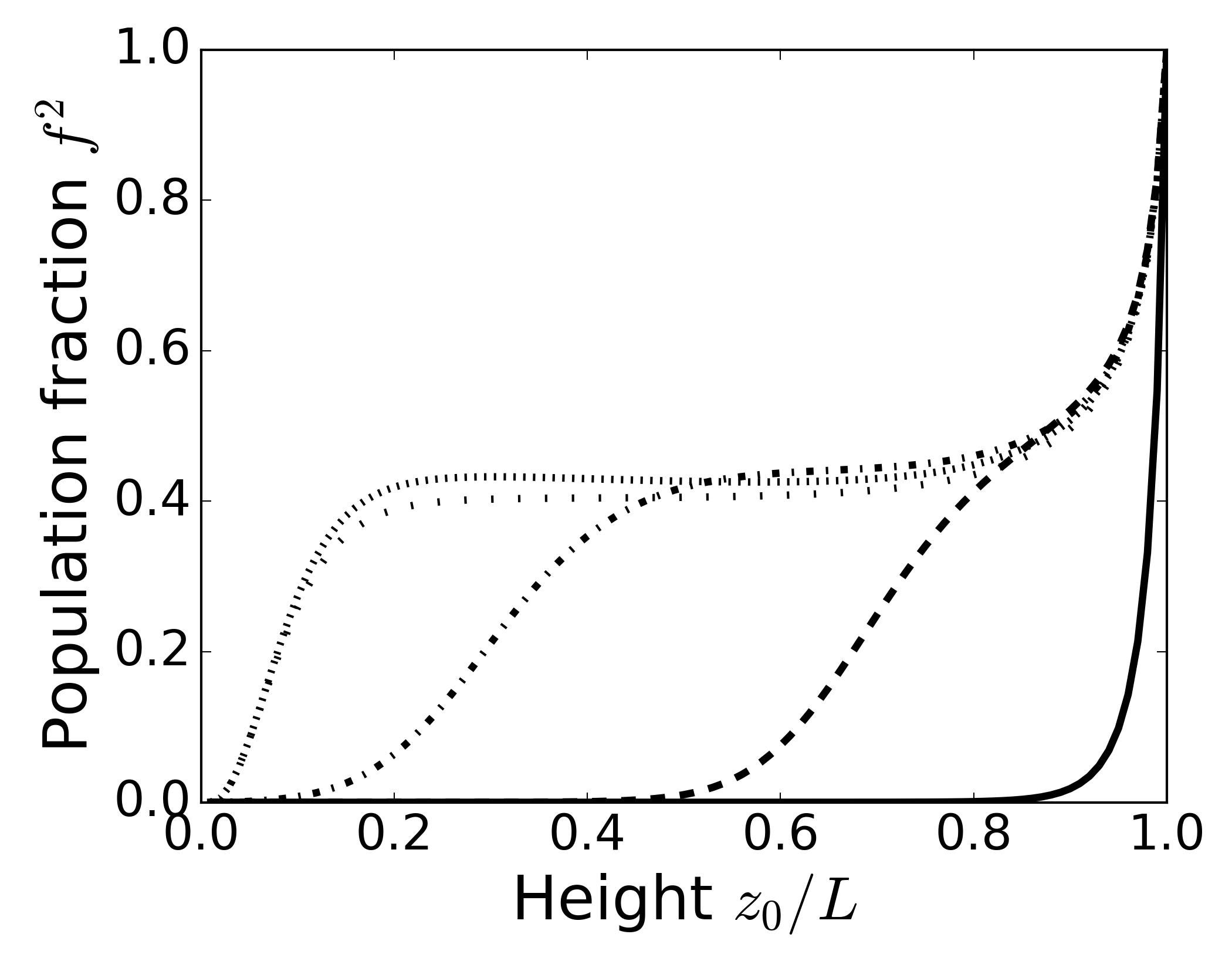}}
    \caption{Fraction of mesogens that respond collectively to the applied electric field $f^2$, obtained from the scaled theory and as a function of the scaled height co-ordinate ${z}_0/L_0$. The results correspond to a super-critical scaled field strength $h=1.48$, with the different curves corresponding to different points in scaled time: $\tilde{t}=0$ (solid), $\tilde{t}=5$ (dashed), $\tilde{t}=10$ (dashdot), $\tilde{t}=15$ (dotted), and $\tilde{t}=60$ (loosely dotted); the profile does not change appreciably afterward. See Sec. \ref{sec:scalings} for the full scaling procedure. The values used for the scaled model parameters are $\zeta=0.51$, $\tilde{B_1}=3.70$, $\eta_\text{ref}=0.675$, $\tilde{\kappa}_f=0.071$, $\tilde{\kappa}_{\tilde{\eta}}=0.071$, $\Gamma_f=0.46$ and $\Gamma_{\tilde{\eta}}=1$ for the kinetic coefficients associated with each order parameter, and $\tilde{\gamma}=0.285$ for the coefficient of viscoelastic relaxation.}\label{fig:traces1}
\end{figure}

\begin{figure}[htbp]
    {\includegraphics[width=8.cm]{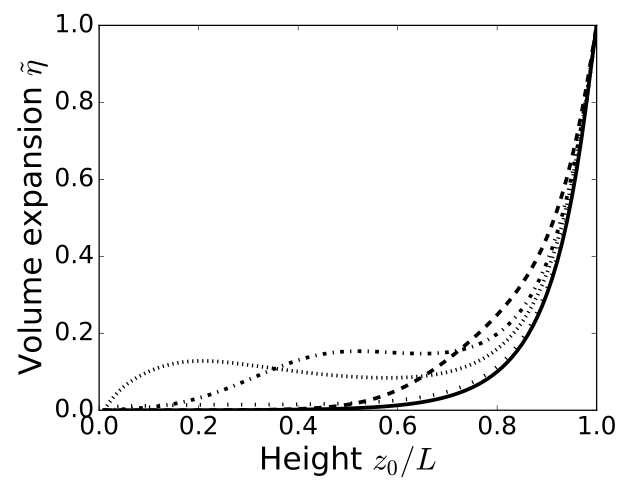}}
    \caption{Relative volume increase of the liquid crystal network $\tilde{\eta}$, obtained from the scaled theory and as a function of the scaled height co-ordinate ${z}_0/L_0$. The results correspond to a super-critical scaled field strength $h=1.48$, with the different curves corresponding to different points in scaled time: $\tilde{t}=0$ (solid), $\tilde{t}=5$ (dashed), $\tilde{t}=10$ (dashdot), $\tilde{t}=15$ (dotted), and $\tilde{t}=60$ (loosely dotted); the profile does not change appreciably afterward. See Sec. \ref{sec:scalings} for the full scaling procedure and see figure \ref{fig:traces1} for the used parameter values.}\label{fig:traces2}
\end{figure}

Figures \ref{fig:traces1} and \ref{fig:traces2} show the time evolution of both order parameter profiles upon application of a homogeneous and super-critical electric field, as follows from the (scaled) relaxational dynamics described in Secs. \ref{sec:dynamics} and \ref{sec:scalings}. The increasingly dotted curves denote different---but equally spaced---points in time.

Figure \ref{fig:traces1} indicates that the response of the population fraction order parameter, $f\left(\tilde{z}_0\right)^2$, to the electric field starts at the top of the film and from there permeates the liquid crystal network. This continues approximately until the dotted curves, corresponding to the time $\tilde{t}=15$, is reached. After this the profile changes very little, as evidenced by the loosely dotted curve, which denotes the asymptotic order parameter profile. This is because only the volume of the liquid crystal network relaxes viscoelastically; the slight relaxation of $f^2$ beyond the dotted curve is entirely due to the coupling to the volume.

This last point is explicitly reflected in figure \ref{fig:traces2}, which shows the response of the volume-expansion order parameter, $\tilde{\eta}\left(\tilde{z}_0\right)$. From this figure we find the same permeation trend, which here assumes the appearance of an inward-moving bulge due to the viscoelastic relaxation of the polymer network. That is, as the front of the response moves deeper into the film, relaxation of the volume elements near the top of the film already commences in its wake. The loosely dotted curve finally indicates that the order parameter profile asymptotically relaxes back to its initial profile.

\begin{figure}[htbp]
    {\includegraphics[width=8.cm]{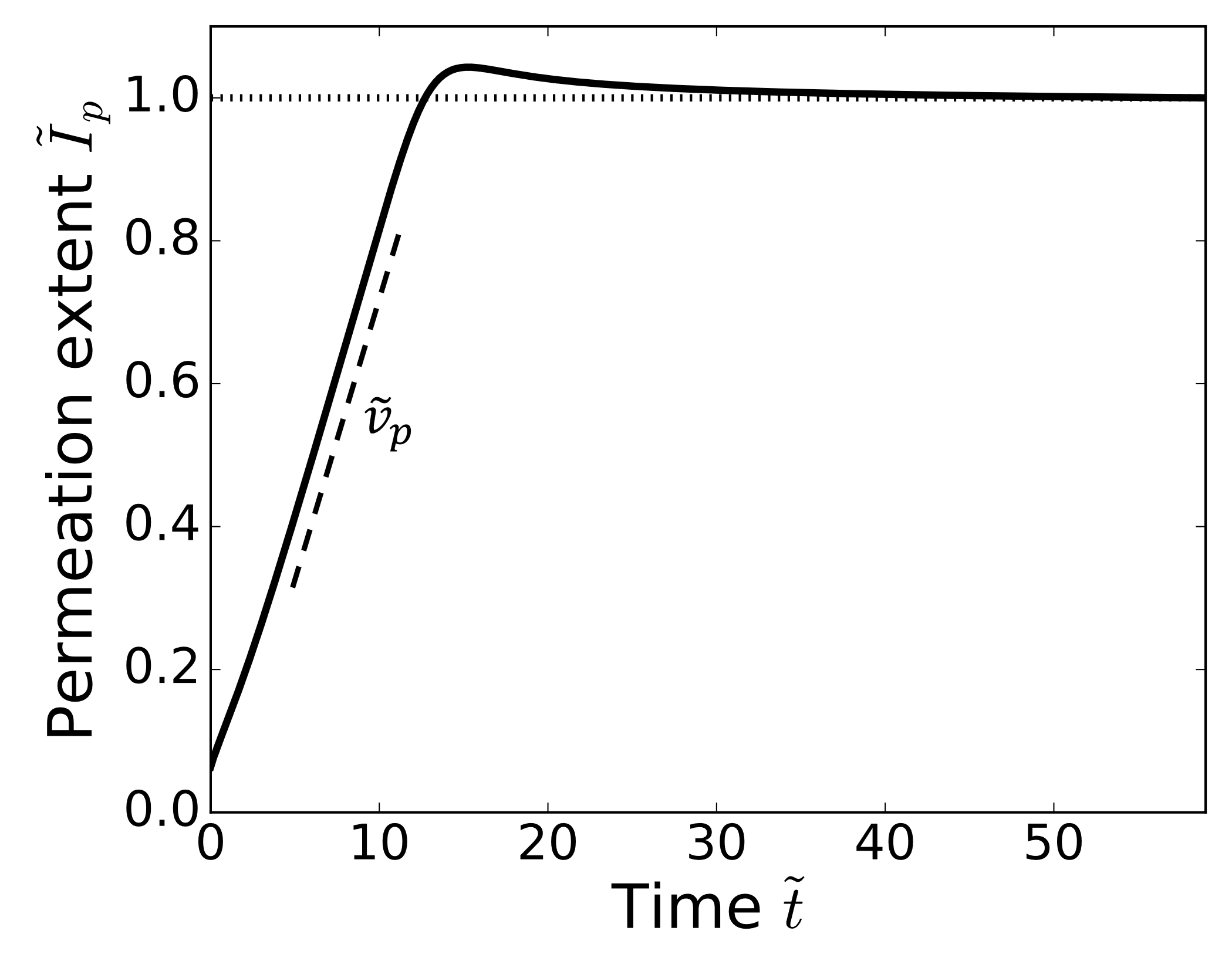}}
    \caption{Permeation extent $\tilde{I}_p$, i.e., how deeply the response permeates the liquid crystal network relative to the final configuration it assumes, as a function of scaled time $\tilde{t}$. The dotted line denotes the asymptotic value $\tilde{I}_p=1$ and the dashed line indicates the constant permeation rate $\tilde{v}_p=\diff \tilde{I}_p/\diff \tilde{t}$. See Sec. \ref{sec:scalings} for the full scaling procedure and see figure \ref{fig:traces1} for the used parameter values.}\label{fig:traces3}
\end{figure}

To quantify the inward motion of the response, we introduce a permeation extent $\tilde{I}_p\left(\tilde{t}\right)$, which measures how deeply the response has penetrated the liquid crystal network at a given time. Anticipating that permeation is dominated by the reorientation of mesogens---we will show this below in Sec. \ref{sec:front}---we do this by measuring the area under the $f^2\left(\tilde{z}_0\right)$ curve in figure \ref{fig:traces1},  relative to its asymptotic profile (loosely dotted curve):
\begin{equation}\label{eq:perm extent}
    \tilde{I}_p\left(\tilde{t}\right)=\frac{\int_0^1\diff \tilde{z}_0 \, f\left(\tilde{z}_0,\tilde{t}\right)^2}{\lim_{\tilde{t}\rightarrow\infty}\int_0^1\diff \tilde{z}_0 \, f\left(\tilde{z}_0,\tilde{t}\right)^2}.
\end{equation}
We cannot use the volume-expansion order parameter profile in the same manner because it fully relaxes viscoelastically. Figure \ref{fig:traces3} shows the corresponding time trace. The response travels inward with an approximately constant velocity, until the order parameters approach their asymptotic profiles. The slight overshoot can be attributed to the viscoelastic relaxation of the polymer network. That is, the volume of the liquid crystal network relaxes viscoelastically, and this results in a slight relaxation of population fraction profile because the two order parameters are coupled. The permeation extent then finally approaches $I_p=1$ (dotted line) because we normalize with respect to the asymptotic profile.

% couples to the population fraction and induces a slight relaxation of its profile, resulting in the found overshoot.

% This motivates us to characterize the permeation behavior by introducing a permeation depth $\tilde{d}_p\left(\tilde{\tau}\right)$, which measures how deeply the response has penetrated the liquid crystal network at a given time \footnote{Specifically, we measure the area underneath the profile of the population fraction $f\left(\tilde{z}_0\right)^2$, relative to the profile it assumes in the limit $\tilde{\tau}\rightarrow\infty$; we cannot use the volume-expansion order parameter profile in the same manner because it fully relaxes viscoelastically. This provides us with an indication of how swiftly the response permeates the liquid crystal network that is mathematically well-defined, according to $\tilde{d}_p\left(\tilde{\tau}\right)=\frac{\int_0^1\diff \tilde{z}_0 \, f\left(\tilde{z}_0,\tilde{\tau}\right)^2}{\lim_{\tilde{\tau}\rightarrow\infty}\int_0^1\diff \tilde{z}_0 \, f\left(\tilde{z}_0,\tilde{\tau}\right)^2}$}.

This establishes the phenomenon of permeation in thin-film liquid crystal networks. Below, we solve the corresponding linearized dynamical equations to estimate how the permeation rate depends on model parameters; we discuss the relevance to experimental parameters in Sec. \ref{sec:interpretations}.

% \subsection{Permeation in terms of front propagation}
\section{Permeation rate}\label{sec:front}
% \textbf{Narratively, this section does something else: interpret the data with a brief bit of theory...}
To estimate the rate at which the response permeates the liquid crystal network, we linearize the dynamical equations about the linearly unstable state $f=\tilde{\eta}=0$ to yield
\begin{equation}\label{eq:LMS}
    \begin{split}
        \partial_{\tilde{t}}f&=h\tilde{\Gamma}_ff+\tilde{\kappa}_f^2\tilde{\Gamma}_f\partial_{\tilde{z}_0}^2f,\\
        \partial_{\tilde{t}}\tilde{\eta}&=-\zeta\tilde{\Gamma}_{\tilde{\eta}}\tilde{\eta}+\tilde{\kappa}_{\tilde{\eta}}^2\tilde{\Gamma}_{\tilde{\eta}}\partial_{\tilde{z}_0}^2\tilde{\eta}-\tilde{\gamma}\left(\tilde{\eta}-\tilde{\eta}_0\right)\tilde{\tau}.
    \end{split}
\end{equation}
The main take-away from Eqn. \eqref{eq:LMS} is that the order parameters decouple in the linearly unstable region into which the response permeates. The dynamics of the population order parameter $f$ takes the form of a diffusion---or, alternatively, heat---equation with added driving term: for super-critical fields ($h>0$) the equation becomes linearly unstable about $f=0$. Conversely, the dynamics of the volume-expansion $\tilde{\eta}$ order parameter only exhibits stabilizing terms, indicating that permeation must be dominated by the population order parameter $f$; the corresponding volume expansion only follows from non-linear couplings that become important deeper within the propagating front.

% Crucially, we can understand the predominantly linear permeation behavior in terms of linear marginal stability theory \cite{van2003front}, which describes the (asymptotic) propagation of a stable front ($f^2,\tilde{\eta}>0$) into a linearly unstable phase ($f^2,\tilde{\eta}=0$). Specifically, for the model we consider the front is ``pulled along" by its tip, justifying a linearization of the dynamical equations about the unstable state $f^2=\tilde{\eta}=0$. Using Eq. \eqref{eq: dynamics}, the result reads \textbf{I do not think I can print these equations if I let readers skip the theory...}
% \begin{equation}\label{eq:LMS}
%     \begin{split}
%         \partial_{\tilde{\tau}}f&=\frac{h-h_*}{27}\tilde{\Gamma}_ff+\tilde{K}_f\tilde{\Gamma}_f\partial_{\tilde{z}_0}^2f,\\
%         \partial_{\tilde{\tau}}\tilde{\eta}&=-\zeta\tilde{\Gamma}_{\tilde{\eta}}\tilde{\eta}-\tilde{\gamma}\tilde{\Gamma}_{\tilde{\eta}}\left(\tilde{\eta}-\tilde{\eta}_0\right)\langle\tilde{\tau}\rangle_{\tilde{\eta}}+\tilde{K}_{\tilde{\eta}}\tilde{\Gamma}_{\tilde{\eta}}\partial_{\tilde{z}_0}^2\tilde{\eta},
%     \end{split}
% \end{equation}
% indicating that near the propagating front tip the order parameters decouple --- note the similarity to the diffusion equation or, alternatively, the heat equation for both. From Eq. \eqref{eq:LMS} we conclude that only the population fraction $f^2$, driven by the electric field $h>h_*$, is linearly unstable; growth of the volume-expansion order parameter $\tilde{\eta}$ only follows from non-linear couplings that become important deeper within the propagating front.

To more closely investigate the dominant permeation dynamics, we solve the dynamical equation of the population order parameter $f$ (first line of equation \eqref{eq:LMS}) by means of a finite Fourier transform; see Appendix \ref{app:FFT} for the details. We then insert the solution into Eqn. \eqref{eq:perm extent} to obtain the permeation extent $\tilde{I}_p$, and compute the permeation rate as $\tilde{v}_p=\diff\tilde{I}_p/\diff\tilde{t}$. In the short-time limit $\tilde{t}\rightarrow0$ the analytical prediction reads
\begin{equation}\label{eq:prediction}
    \tilde{v}_p=\frac{1+h}{2}\tilde{\Gamma}_f\frac{\tilde{\kappa}_f\sinh{2/\tilde{\kappa}_f}-2}{\sinh^2{1/\tilde{\kappa}_f}}.
\end{equation}
We shall return to the experimental implications of this expression in Sec. \ref{sec:interpretations}. First, we test the accuracy of this prediction by comparing it to the permeation rate exhibited by the full theory. This we measure numerically by means of a linear fit to the permeation extent $\tilde{I}_p$ for a range of field strengths $h$ and square gradient coefficients $\kappa_f$. We restrict the range of the fit to the regime of linear growth, such that the slope yields the average permeation rate (see Fig. \ref{fig:traces3}). 
This is to be compared with the analytical prediction of Eqn. \eqref{eq:prediction}, applicable to the short-time limit.

% Based on equation \eqref{eq:LMS}, the theory of linear marginal stability predicts a rate of permeation $\tilde{v}_p=2\tilde{\Gamma}_f\tilde{\kappa}_f\sqrt{h}$. We shall return to the experimental implications of this expression in Sec. \ref{sec:interpretations}. First, we test the accuracy of this prediction by comparing it to the permeation rate exhibited by the full theory. This we measure numerically by means of a linear fit to the permeation extent $\tilde{I}_p$ for a range of field strengths $h$ and square gradient coefficients $\kappa_f$.

\begin{figure}[htbp]
    {\includegraphics[width=8.cm]{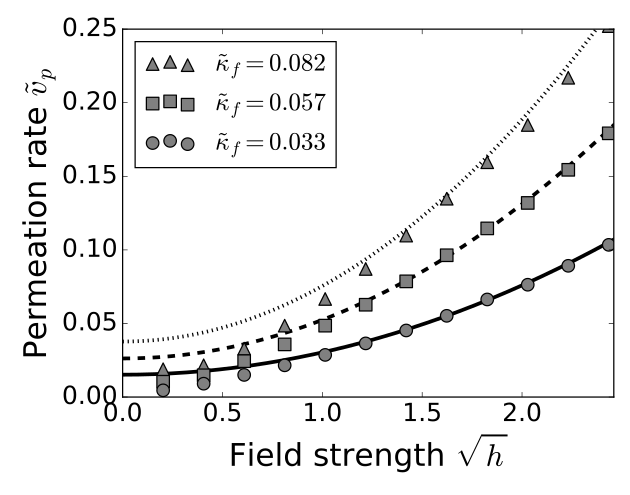}}
    \caption{Permeation rate as a function of the scaled field strength $\sqrt{h}$. Symbols (circle, square, triangle) represent data points obtained from the scaled theory and lines (solid, dashed, dotted) represent the short-time prediction of Eqn. \eqref{eq:prediction} (scaled identically); different symbols/lines correspond to different values of the scaled square gradient coefficient $\tilde{\kappa}_f$. See Sec. \ref{sec:scalings} for the full scaling procedure and see figure \ref{fig:traces1} for the used parameter values not explicitly stated here.}\label{fig:permeation1}
\end{figure}

Figure \ref{fig:permeation1} shows the permeation rate as a function of the field strength $h$ for various values of $\tilde{\kappa}_f$ (different symbols), as compared with the short-time prediction described above (different curves). We generally find good agreement, although the prediction does not perform as well for low electric field strengths. The prediction does not approach zero as $h\rightarrow0$ because $h=0$ only coincides with the critical field strength in the bulk of the film, where $f=\tilde{\eta}=0$; near the free surface of the film the critical field strength is smaller, implying that permeation can still commence. The discrepancy with the full theory follows from the permeation no longer being strictly linear in time for sufficiently weak electric fields, meaning our short-time approximation breaks down. 

\begin{figure}[htbp]
    {\includegraphics[width=8.cm]{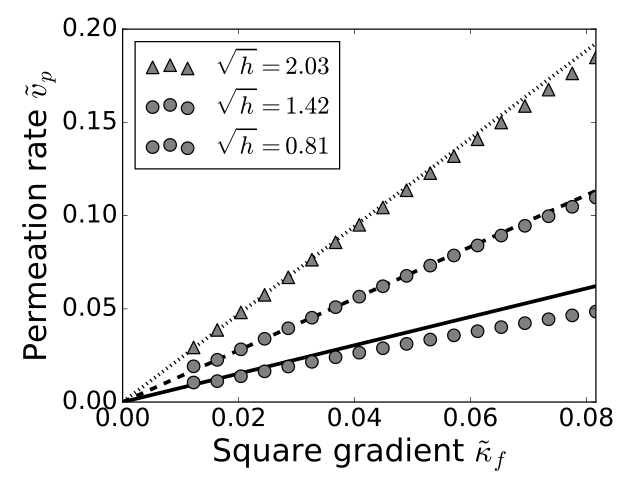}} \\
    \caption{Permeation rate as a function of the scaled square gradient coefficient $\tilde{\kappa}_f$. Symbols (circle, square, triangle) represent data points obtained from the scaled theory and lines (solid, dashed, dotted) represent the short-time prediction of Eqn. \eqref{eq:prediction} (scaled identically); different symbols/lines correspond to different values of the scaled field strength $\sqrt{h}$. See Sec. \ref{sec:scalings} for the full scaling procedure and see figure \ref{fig:traces1} for the used parameter values not explicitly stated here.}\label{fig:permeation22}
\end{figure}

Figure \ref{fig:permeation22} shows the variation of the permeation rate with the square gradient coefficient $\tilde{\kappa}_f$, for various values of the electric field strength. Here we again find that short-time prediction generally agrees well with the full theory, with the prediction performing somewhat worse for weak electric field strengths; this is in accordance with figure \ref{fig:permeation1}. We do not show more data near $\tilde{v}_p=0$ because numerically solving the dynamical equations for a sufficiently long time $\tilde{t}$ and with sufficient accuracy becomes progressively more demanding in this regime.

Thus, we have established permeation as a mechanism in thin-film liquid crystal networks and derived a relatively simple closed-form expression for the corresponding permeation rate, which compares favorably with our full theory. Below, we take a step back and discuss the possible interpretations our theory admits, as well as their relation to experimentally-accessible parameters.

% Below, we summarize the most salient findings of this paper as a whole and discuss their significance, possible interpretations, and their relation to experimentally-accessible parameters.

\section{Interpretations}\label{sec:interpretations}

At the start of this paper we pointed out that although the theory we present here was originally derived based on excluded-volume arguments, this is no longer explicitly reflected in the equations we show. This makes the theory, which we previously showed to qualitatively agree with both molecular dynamics computer simulations and experiments \cite{kusters2020dynamical}, increasingly versatile, as it also admits alternative interpretations of the underlying mechanisms without (significant) alterations to the theory. Here, we focus on interpretations based on (1) the aforementioned excluded-volume arguments, (2) an electrically-driven glass transition, and (3) the cooperative reorientation of an array of oscillators, which we discuss in order below.

If we derive the theory from excluded-volume arguments, hard-core mesogen-mesogen interactions dominate the response of the liquid crystal network. Onsager showed that for two rod-like particles of length $l$ and diameter $d$, these interactions scale as $l^2d\lvert\sin\varphi\rvert$, with $\varphi$ the angle between their axes \cite{onsager1949effects}. This suggests that both the aspect ratio of the mesogens, as well as the degree of orientational order with which they are cross-linked into the network, are important experimental parameters. Within our model, they specifically contribute to the critical field strength required to locally align dipolar mesogens with the electric field, i.e., to overcome their excluded-volume interactions with neighboring cross-linked mesogens. Since the square gradient contributions to the theory are derived from the same local interactions, the coefficient $\tilde{\kappa}_f$ is expected to also be an increasing function of said experimental parameters.

Connecting this to our prediction of the permeation rate, Eqn. \eqref{eq:prediction}, which is arguably the most salient finding of this paper, we conclude that the aspect ratio of the mesogens and the degree of orientational order with which they are cross-linked into the network can be tuned to control permeation. Specifically, the field strength $h$, which we scaled to its critical value, is a decreasing function of these parameters, whereas the square gradient coefficient $\tilde{\kappa}_f$ is an increasing function of them, leading us to expect that variation of either experimental parameter gives rise to a single optimum.

Alternatively, recent experimental work of Van der Kooij and co-workers suggests that an electrically-driven glass transition could lie at the root of the electrically-deforming liquid crystal networks under consideration \cite{van2020electroplasticization}. This notion is further corroborated by computer simulations on kinetically-facilitated Ising models of glass-forming liquids, which show a linear permeation phenomenon similar to ours \cite{leonard2010macroscopic,tito2015enhanced}. The underlying rationale is that dipolar mesogens inside the network are locally ``caged" by the neighboring mesogens. Upon actuation, these dipolar mesogens cannot fully escape their cages---all mesogens are connected to the polymer network---but they can effectively enlarge their cages by means of collective reorientation. Once a sufficient fraction of mesogens (both dipolar and crosslinked) reorients collectively, an appreciable number of cages becomes enlarged, potentially weakening the polymer matrix and causing volume expansion.

% In addition, it should be noted that computer simulations on kinetically facilitated Ising models  \cite{leonard2010macroscopic,tito2015enhanced}

% In this case it is expected that appreciable volume changes ensue once a sufficient fraction of mesogens (both dipolar and crosslinked) reorients collectively. 

In terms of theory, Gutzow and co-workers previously proposed a Landau-type theory, of the Ising universality class, to phenomenologically describe the glass transition \cite{gutzow2004glass,gutzow2010glass}. In their work, the relevant order parameter was the difference in molar entropy between the glassy and the rubbery state. Our theory exhibits a similar form for the population fraction order parameter $f^2$, which we subsequently couple to the volume expansion of the liquid crystal network. Although the above provides a generic point of connection for our theory in the context of a glass transition, in order to link to experimentally-accessible parameters we must make explicit reference to an underlying mechanism. For example, if we assume the coupling between two arbitrary mesogens is primarily mediated by other mesogens, with which the network is very densely packed, we again conclude that, in addition to the electric field strength, the aspect ratio of the mesogens and the degree of orientational order with which they are cross-linked into the network govern the permeation rate.

The above can be made more concrete by recognizing that, apparently, couplings between different mesogens, mediated by some matrix, are important. One way to make such a coupling explicit is by treating the mesogens as a collection of interacting oscillators, which is the final interpretation we discuss here. For simplicity, we assume the oscillators obey the Kuramoto model, which states that all oscillators interact with each other with a coupling strength proportional to the difference in their phase angles \cite{acebron2005kuramoto}; this model contains no spatial information. It can be shown that this essentially constitutes a mean-field model identical in form to the theory for the population fraction $f^2$ we describe in Sec. \ref{sec:Theory} \cite{daido1994generic,daido1996onset}. If we reason from this viewpoint, $f^2$ could, alternatively, be interpreted as the fraction of mesogens that oscillate collectively, while the ``field strength" $h$ represents an effective coupling constant scaled to its critical value. The latter is not unreasonable, since stronger fields exert larger torques on the dipolar mesogens, thus increasing the deflections they make.

Interpreting the field in terms of a coupling constant also suggests a dependence on the aspect ratio of the mesogens and  the degree of orientational order with which they are cross-linked into the network, both of which increase the likelihood of a reorienting mesogen encountering another mesogen on its path. Thus, we again identify said parameters as the most important in governing permeation, in addition to the electric field strength itself. The exact form of this relationship depends on whether the coupling constant grows more strongly with these parameters than its critical value. If so, we recover a permeation rate that monotonically increases with the mentioned experimental parameters, in addition to the (actual) electric field strength, whereas we recover a single optimum upon varying the aspect ratio of the mesogens and their orientational order crosslinked into the network otherwise.

\section{Conclusion \& discussion}\label{sec:conclusion}
% \textbf{Beetje dubbelop met vorige alinea...}

In summary, we have developed a spatially-inhomogeneous theory for electrically-deforming liquid crystal networks. We applied it to the expansion of thin liquid crystal network films, for which the theory predicts that if the system is actuated homogeneously expansion starts at the free surface of the film, and subsequently permeates the liquid crystal network film; the experimental reality can be somewhat more complicated, to which we return below. This finding is significant, since it allows us to directly probe how the response timescales of these materials can be controlled in experimentally-relevant geometries. We showed that permeation generally occurs at a constant rate and we provided a closed-form prediction of the permeation rate in terms of experimentally-accessible model parameters. This suggests permeation can be controlled by varying the aspect ratio of the mesogens and their degree of orientational order when cross-linked into the polymer network. Specifically, we predict variation of either parameter to give rise to a single optimum in the permeation rate. Exploiting this insight could pave the way for the controlled release, retention and exchange of molecular cargo.

The permeation phenomenon we report on here can be directly verified and exploited using currently available experimental systems. Light-driven liquid crystal networks, containing azobenzene moieties that undergo a \textit{trans-to-cis} interconversion upon illumination with suitable wavelength, provide a prime example of this, since these are generally illuminated homogeneously from above. The electrically-actuated systems that are the focus of this paper, however, are usually driven by spatially-\textit{inhomogeneous} fields that are significantly weaker near the top of the film \cite{liu2017protruding,van2019morphing,van2020electroplasticization}. In this case the theory predicts no clear top-to-bottom permeation, suggesting permeation can only be realized if the experimental design is adapted.
To reinforce this recommendation, we also conducted preliminary Molecular Dynamics computer simulations that suggest, in accordance with the theory, that permeation \textit{does} occur in homogeneously-actuated electrically-deforming liquid crystal networks (see the Supplementary Video) \footnote{The simulations are carried out using the HOOMD-blue simulation toolkit. See Refs \cite{liu2017protruding} and \cite{kusters2020dynamical} for more details on the simulation procedure, which we extended to a thin-film geometry for the purpose of our preliminary study}; we intend to discuss and expand upon these findings in a forthcoming publication \cite{futurework2}.

The above implies that permeation in liquid crystal networks can be accessed by applying different electric field geometries or protocols, such as a spatially-homogeneous field, which could pave the way for a finer control over the motion of molecular cargo, e.g., for release, uptake and transport.
Our findings pertaining to permeation also have broader implications for the time scales on which a liquid crystal network film macroscopically deforms. This should not surprise us, as the rate of permeation is related to the rate at which the film thickness increases before it viscoelastically relaxes in order to assume its original shape. Accordingly, the time required to achieve maximum volume expansion, the so-called rise time $\tilde{t}_r$, must scale inversely with the permeation rate (see Appendix \ref{app:rise time}).

In addition, it is worth pointing out that the experimental parameters we identify as governing permeation are---at least partially---the same as those we previously showed to dictate the \textit{magnitude} of volume expansion \cite{kusters2020dynamical}. Although, naively, this may lead us to suspect both can be optimized simultaneously in applications, we have carried out exploratory numerical tests on AC-driven liquid crystal network films that suggest this is not the case. That is, the time scales optimizing how deeply the response to an alternating electric field permeates the liquid crystal network---in the steady state---are distinct from those optimizing its overall expansion. This means a trade-off must be made, similar to how the thickness of the film balances the response magnitude and the response time, suggesting specialized applications likely require different chemical ingredients to function optimally.

Finally, we remind the reader that, in the interest of presenting a digestible narrative, we opted for a simplified boundary condition in this paper. This filtered out some of the additional model features that are of secondary importance to the permeation phenomenon we highlight here. The most striking of these features is the existence of a ``wait time", after applying the electric field, before appreciable volume expansion is observed. This coincides with the time required for a significant population of collectively-responding mesogens to form; a similar ``wait time" has been observed experimentally upon application of an alternating electric field \cite{van2019morphing}. We previously discussed this feature in the context of the bulk theory, in Ref. \cite{kusters2020dynamical}. Our simplified boundary condition bypasses this feature by imposing a non-zero population at the free surface of the film, where we stress that our findings in this paper remain valid even if this condition is relaxed.

% We previously discussed the above feature in the context of the bulk theory in Ref. \cite{kusters2020dynamical}, where we stressed its relevance to experiments ... 

% ... metastable initial condition if field turned on, exponential growth ... link to experiments ...

% ... metastable initial condition paves way for bulk-like response ... 

% ... Randconditie: dit is lastig, aangezien ik weinig heb om op terug te vallen wat betreft verwijzing ...

\section{Outlook}\label{sec:outlook}
Next, we discuss alternative approaches through which the response of the liquid crystal network to external fields can be controlled. Although the focus of this work is on analyzing the phenomenon of permeation itself, for which constant electric fields are the most suitable, in a follow-up publication we intend to extend our analysis to alternating electric fields, as are also used in experiments \cite{liu2017protruding,van2019morphing,van2020electroplasticization}. The main advantage of this approach is that the response can be sustained, and preliminary calculations even suggest the response can be spatially contained and thus steered \cite{futurework}.

In addition, we may consider different geometries. In this paper, we focused on a thin film clamped to a substrate, but we may just as easily consider a film that freely expands in either direction. In that case, the expansion of the film is no longer hampered by the substrate, and can permeate inward from both free surfaces. It can be shown that this leads to a greater overall deformation, as well as faster permeation, providing an additional route to controlled actuation.

Another factor we did not consider in detail is the initial state of the liquid crystal network: we assumed strong (uniaxial) orientational order of the mesogens, perpendicular to the electric field axis, throughout this paper. If, instead, we initialize the system in an orientationally \textit{disordered} state, reasoning from the principle of excluded-volume generation we would expect a markedly less-pronounced response. This is because the disordered state already exhibits a high degree of mutually-excluded volume between the mesogens. However, if a purely dynamical transition lies at the root of how these materials deform, a significant response can still be recovered. Thus, by varying the initial configuration of the liquid crystal network, more light can be shed on the mechanism underlying their deformation. In this context, the cholesteric phase is also of some interest, as this presents an additional degree of ordering that must be broken by the external stimulus \cite{feng2018oscillating}.

% In addition, we could vary thickness of the film, which presents a trade-off between the response time and the response magnitude. 

Looking further afield, an obvious extension of the current theory is from one to two spatial dimensions. The implementation of such a model is straightforward, following along the same lines as discussed in Sec. \ref{sec:Theory2}, and can be used to study the emergence of patterns on the film surface. For example, by applying an electric field that moves along the surface of the liquid crystal network, e.g., by translating the electrodes underneath the substrate, we obtain a volume-expansion response resembling a traveling wave on the film surface. Actuation protocols such as this provide an avenue for moving molecular cargo on top of the liquid crystal network surface in a controlled manner.

Finally, although the focus of this paper was on controlling permeation in liquid crystal network films---taking steps toward realizing control over the transport of molecular cargo---our findings also lend themselves to other applications. For example, the inhomogeneous swelling of the liquid crystal network can be utilized in actuatable devices such as artificial muscles or microswimmers, both of which have already been proposed or studied in the context of liquid crystal elastomers \cite{wermter2001liquid,ikeda2007photomechanics,sanchez2009photo,sanchez2011liquid,dai2013humidity,schuhladen2014iris,iamsaard2016fluorinated,zeng2017self,lucantonio2017coupled}. Though it is clear that such applications require a more in-depth consideration of the geometry, actuation protocol and the different strains involved, they do underscore the versatility of the theory we present here, as well as that of the materials under consideration.

\begin{acknowledgments}
	This research received funding from the Dutch Research Council (NWO) in the context of the Soft Advanced Materials (SAM) consortium in the framework of the ENW PPP Fund for the top sectors and from the Ministry of Economic Affairs in the framework of the `PPS-Toeslagregeling'.
\end{acknowledgments}

\appendix
\section{Length scales}\label{app:lengthscales}
To contextualize the relation between the square gradient coefficients, $\tilde{\kappa}_f$ and $\tilde{\kappa}_{\tilde{\eta}}$, and the boundary layers shown in figure \ref{fig:illustrate}, we estimate the corresponding length scales. To this end, we expand the free-energy-functional Eq. \eqref{eq:Gibbsreference} up to second order about $f^2=\tilde{\eta}=0$, which is justified in the bulk of the film as long as we operate below the critical field strength ($h<0$). This yields a linearized equation of state for both order parameter profiles. The solutions are the hyperbolic sines
\begin{equation}
    \begin{split}
        f\left(\tilde{z}_0\right)&=\frac{\sinh{\tilde{z}_0/\tilde{\kappa}_f}}{\sinh{1/\tilde{\kappa}_f}}\\
        \tilde{\eta}\left(\tilde{z}_0\right)&=\frac{\sinh{\tilde{z}_0\sqrt{\zeta}/\tilde{\kappa}_{\tilde{\eta}}}}{\sinh{\sqrt{\zeta}/\tilde{\kappa}_{\tilde{\eta}}}},
    \end{split}
\end{equation}
from which the corresponding length scales $\tilde{\kappa}_f$ and $\tilde{\kappa}_{\tilde{\eta}}/\sqrt{\zeta}$ can be read off straightforwardly. Inserting the parameter values used in the main text, assuming no applied field ($h=-1$), we find $\tilde{\kappa}_f=0.071$ and $\tilde{\kappa}_{\tilde{\eta}}/\sqrt{\zeta}=0.099$, i.e., the boundary layers are expected to extend for $7-10\%$ of the initial (undeformed) film thickness. Clearly, this simple analysis breaks down close to the surface of the film, as well as for super-critical field strengths $h>0$, since non-linear couplings become important in these cases. However, it does provide a frame of reference for interpreting the square gradient coefficients, which distinguish the spatially-inhomogeneous theory from its bulk counterpart, at a base level.

% Note that although the values used for the square gradient coefficients $\tilde{K}_f$ and $\tilde{K}_{\tilde{\eta}}$ in figure \ref{fig:illustrate} may seem arbitrarily small, the order parameter profiles already suggest the length scales associated with them can be significant on the scale of the thin film. To contextualize this, we provide a rough estimate of said length scales by expanding the free-energy-functional Eq. \eqref{eq:Gibbsreference} up to second order about $f^2=\tilde{\eta}=0$, which is justified in the bulk of the film as long as we operate below the critical field strength ($h<0$). This yields a linearized equation of state for both order parameter profiles, which we solve by means of a finite Fourier transform. The resulting profiles take the form of hyperbolic sines, where the arguments are scaled with the length scales $l_f/L_0=\tilde{\kappa}_f/\sqrt{-h}$ and $l_{\tilde{\eta}}/L_0=\tilde{\kappa}_\eta/\sqrt{\zeta}$ for the respective order parameters. Inserting the parameter values used above, assuming no applied field ($h=-1$), we find $l_f/L_0=0.071$ and $l_{\tilde{\eta}}/L_0=0.10$. Clearly, this simple analysis breaks down close to the surface of the film, as well as for super-critical field strengths $h>0$, since non-linear couplings become important in these cases. However, it does provide a frame of reference for interpreting the square gradient coefficients, which distinguish the non-local theory from its local counterpart, at a base level.

\section{Finite Fourier transform}\label{app:FFT}
In this appendix we solve the linearized dynamical equation for the order parameter $f$ by means of a finite Fourier transform. The starting point of this analysis is given by Eq. \eqref{eq:LMS}, which is reprinted here for reference:
\begin{equation}\label{eq:A1}
    \partial_{\tilde{t}}f=h\,\tilde{\Gamma}_ff+\tilde{\kappa}_f^2\tilde{\Gamma}_f\partial_{\tilde{z}_0}^2f.
\end{equation}
Before transforming Eq. \eqref{eq:A1} we find it convenient to first substitute $\varphi=f-\tilde{z}_0$, making the boundary conditions become homogeneous, $\varphi\left(0,\tilde{t}\right)=0=\varphi\left(1,\tilde{t}\right)$. The corresponding initial condition is obtained by solving Eq. \eqref{eq:A1} in the absence of an electric field, giving $\varphi\left(\tilde{z}_0,0\right)=\frac{\sinh{\tilde{z}_0/\tilde{\kappa}}}{\sinh{1/\tilde{\kappa}}}-\tilde{z}_0$.

Next, we introduce the finite Fourier transform as
\begin{equation}\label{eq:a2}
    \overline{\varphi}_n\left(\tilde{t}\right)=\int_0^1\diff \tilde{z}_0 \, \varphi\left(\tilde{z}_0,\tilde{t}\right)\sqrt{2}\sin{n\pi\tilde{z}_0}, \, \, \, \, \, \, \, \, \, \,  n\in \mathbb{N}^+,
\end{equation}
which transforms Eq. \eqref{eq:A1} to
\begin{equation}
    \partial_{\tilde{t}}\overline{\varphi}_n=\left(h-n^2\pi^2\tilde{\kappa}_f^2\right)\tilde{\Gamma}_f\overline{\varphi}_n+\frac{\left(-1\right)^{n+1}\sqrt{2}\,h \, \tilde{\Gamma}_f}{n\pi},
\end{equation}
where the boundary conditions are satisfied by construction. The transformed equation can be solved straightforwardly for $\overline{\varphi}_n$, taking into account the appropriate initial condition, to yield
\begin{equation}
    \overline{\varphi}_n\left(\tilde{t}\right)=\frac{\left(-1\right)^n\sqrt{2}\left[\left(1+h\right)n^2\pi^2\tilde{\kappa}_f^2e^{\left(h-n^2\pi^2\tilde{\kappa}_f^2\right)\tilde{\Gamma}_f\tilde{t}}-h\left(1+n^2\pi^2\tilde{\kappa}_f^2\right)\right]}{n\pi\left(1+n^2\pi^2\tilde{\kappa}_f^2\right)\left(-h+n^2\pi^2\tilde{\kappa}_f^2\right)}.
\end{equation}
This is finally mapped back to real space by summing over all Fourier modes, according to
\begin{equation}\label{eq:a5}
    f\left(\tilde{z}_0,\tilde{t}\right)=\tilde{z}_0+\sum_{n=1}^\infty\overline{\varphi}_n\left(\tilde{t}\right)\sqrt{2}\sin{n\pi\tilde{z}_0}.
\end{equation}

\begin{figure}[htbp]
    \subfloat[]{\includegraphics[width=8.cm]{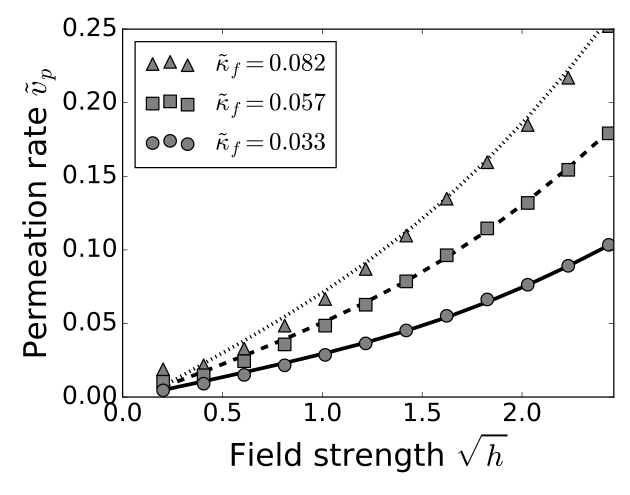}}
    \subfloat[]{\includegraphics[width=8.cm]{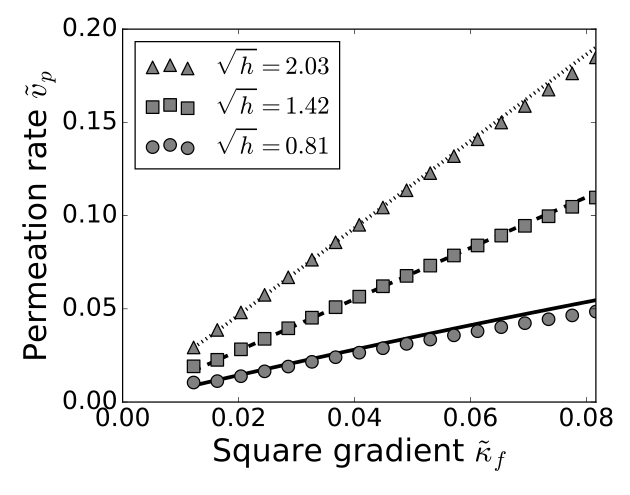}} \\
    \caption{Permeation rate as a function of the scaled field strength $\sqrt{h}$ (a) and the scaled square gradient coefficient $\tilde{\kappa}_f$ (b). Symbols (circle, square, triangle) represent data points obtained from the scaled theory and lines (solid, dashed, dotted) represent the corresponding finite Fourier solution; different symbols/lines correspond to different values of the scaled square gradient coefficient $\tilde{\kappa}_f$ (a) and the scaled field strength $\sqrt{h}$ (b). See Sec. \ref{sec:scalings} for the full scaling procedure and see figure \ref{fig:traces1} for the used parameter values. To be compared with figures \ref{fig:permeation1} and \ref{fig:permeation22} in the main text.}\label{fig:permeation2}
\end{figure}

We test the accuracy of the solution by computing the permeation rate according to the procedure described in the main text. Figure \ref{fig:permeation2} shows the results as a function of the electric field strength $h$ (a) and the square gradient coefficient $\tilde{\kappa}_f$ (b). Symbols denote the results obtained from the full scaled theory, whereas the lines correspond to the permeation rate obtained based on Eqn. \eqref{eq:a5} (incorporating the first 1000 terms in the expansion). The good agreement apparent from figure \ref{fig:permeation2} suggests that non-linear coupling, which are neglected in this approach, are of secondary importance in determining the permeation rate.

Although Eqn. \eqref{eq:a5} accurately predicts the permeation rate, it is not a tractable solution: we must still determine the order parameter profile numerically by summing over all Fourier modes to obtain the solution. To address this, we seek an approximation that can provide a closed-form prediction of the permeation rate in the short-time limit. 

Since we measure the permeation rate by integrating the area under the $f^2\left(\tilde{z}_0\right)$ curve, we compute the integral
\begin{equation}
    \int\diff\tilde{z}_0 \, f^2\left(\tilde{z}_0\right)=\frac{1}{3}+\sum_{n=1}^\infty\left[\frac{2\sqrt{2}\left(-1\right)^{n+1}}{n\pi}\overline{\varphi}_n\left(\tilde{t}\right)+\overline{\varphi}_n^2\left(\tilde{t}\right)\right]
\end{equation}
to this end. 
Subsequently taking the derivative with respect to time and setting $\tilde{t}=0$---this last step amounts to a linearization of the integral in time---we recover
\begin{equation}
    \tilde{v}_p=\sum_{n=1}^\infty\frac{4\left(1+h\right)n^2\pi^2\tilde{\kappa}_f^4\tilde{\Gamma}_f}{\left(1+n^2\pi^2\tilde{\kappa}_f^2\right)}=\frac{1+h}{2}\tilde{\Gamma}_f\frac{\tilde{\kappa}_f\sinh{2/\tilde{\kappa}_f}-2}{\sinh^2{1/\tilde{\kappa}_f}}.
\end{equation}
This is the short-time approximation that we discuss in the main text of the paper.

\section{Rise time}\label{app:rise time}
\begin{figure}[htbp]
    {\includegraphics[width=8.cm]{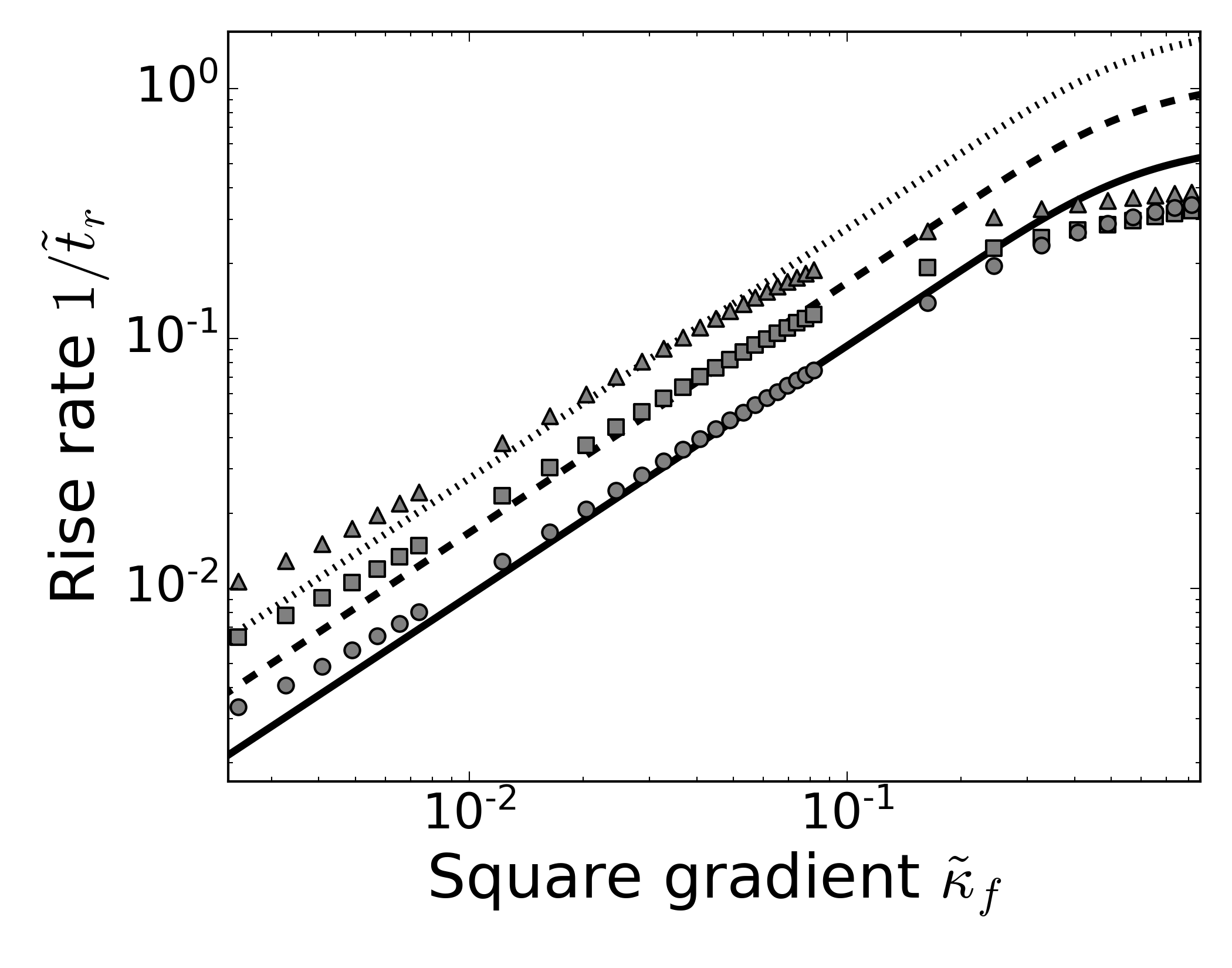}}\\
    \caption{Rise rate of the liquid crystal network film obtained from the scaled theory as a function of the scaled square gradient coefficient $\tilde{\kappa}_f$ (symbols) and the corresponding permeation rate as per Eqn. \eqref{eq:prediction} (lines). We show results for various scaled electric field strengths $h=1.03$ (circles, solid curve), $h=2.63$ (squares, dashed curve) and $h=4.98$ (triangles, dotted curve). See Sec. \ref{sec:scalings} for the full scaling procedure and see figure \ref{fig:traces1} for the used parameter values not explicitly stated here.}\label{fig:rise rate}
\end{figure}

We illustrate the relation between the rise rate and the permeation rate by means of figure \ref{fig:rise rate}, which shows a double-logarithmic plot of the rise rate, i.e., the time required to achieve maximum volume expansion $1/\tilde{t}_r$, as a function of the square-gradient coefficient $\tilde{\kappa}_f$ for various electric field strengths (symbols). The corresponding permeation rates, predicted by Eqn. \eqref{eq:prediction}, are given by the plotted lines (dotted, dashed and drawn) for different field strengths. From this we conclude that there exists a broad range of parameter values for which the rise rate approximately exhibits the expected relationship $1/\tilde{t}_r\approx\tilde{v}_p$; in the limits of low and high $\tilde{\kappa}_f$ the permeation rate is no longer constant and so the linear short-time approximation, which essentially amounts to a linearization, breaks down. Even in the region in figure \ref{fig:rise rate} where the correspondence is relatively good, the permeation rate shows a stronger dependence on $\tilde{\kappa}_f$ than the rise rate due to the viscoelastic relaxation of the polymer network, which becomes comparatively more important as spatial gradients are penalized more strongly. This holds for both $\tilde{\kappa}_f$ and $\tilde{\kappa}_{\tilde{\eta}}$. To support this point we refer to figure \ref{fig:phase diagram f} in the main text, from which we deduce that the volume expansion of the film \textit{relative to its initial configuration} is a decreasing function of $\tilde{\kappa}_f$; the same is true for $\tilde{\kappa}_{\tilde{\eta}}$. 

Finally, we remark that similar agreement between the predicted permeation rate and the rise rate is found upon varying other model parameters, corroborating the close relationship between permeation and macroscopic deformation. Our findings suggest that the rise time of the liquid crystal network film, i.e., the time scale required to ``activate" the material can be controlled through the same experimental parameters that govern permeation.

\bibliography{IDLT_LCN}

\end{document}